\documentclass[english,aip,pop,reprint,floatfix]{revtex4-2}
\usepackage{ae,aecompl}

\usepackage[T1]{fontenc}
\usepackage{amsmath}
\usepackage{amssymb}
\usepackage{graphicx}

\usepackage{xcolor}
\usepackage[normalem]{ulem} 
\usepackage{soul} 

\makeatletter
\DeclareMathOperator\sign{sgn}

\makeatother

\usepackage{babel}
\begin{document}
\title{Numerical study of $\delta$-function current sheets arising from resonant
magnetic perturbations}

\author{Yi-Min Huang}
\email{yiminh@princeton.edu}

\affiliation{Department of Astrophysical Sciences, Princeton University, Princeton, New Jersey 08544, USA}
\affiliation{Princeton Plasma Physics Laboratory, Princeton, New Jersey 08543,
USA}
\author{Stuart R.~Hudson}
\affiliation{Princeton Plasma Physics Laboratory, Princeton, New Jersey 08543,
USA}
\author{Joaquim Loizu}
\affiliation{École Polytechnique Fédérale de Lausanne, Swiss Plasma Center, CH-1015
Lausanne, Switzerland }
\author{Yao Zhou}
\affiliation{Princeton Plasma Physics Laboratory, Princeton, New Jersey 08543,
USA}
\affiliation{Institute of Natural Sciences, School of Physics and Astronomy, and MOE-LSC, Shanghai Jiao Tong University, Shanghai 200240, China}
\author{Amitava Bhattacharjee}
\affiliation{Department of Astrophysical Sciences, Princeton University, Princeton, New Jersey 08544, USA}
\affiliation{Princeton Plasma Physics Laboratory, Princeton, New Jersey 08543,
USA}

\begin{abstract}
General three-dimensional toroidal ideal magnetohydrodynamic equilibria
with a continuum of nested flux surfaces are susceptible to forming singular current sheets when resonant perturbations are applied.
The presence of singular current sheets indicates that, in the presence of non-zero resistivity, magnetic reconnection will ensue, leading to the formation of magnetic islands and potentially regions of stochastic field lines when islands overlap. Numerically resolving singular current sheets in the ideal MHD limit has been a significant challenge. This work presents numerical solutions of the Hahm-Kulsrud-Taylor (HKT) problem, which is a prototype for resonant singular current sheet formation. The HKT problem is solved by two codes: a Grad-Shafranov (GS) solver and the SPEC code. The GS solver has built-in nested flux surfaces with prescribed magnetic fluxes. The SPEC code implements multi-region relaxed magnetohydrodynamics (MRxMHD), whereby the solution relaxes to a Taylor state in each region while maintaining force balance across the interfaces between regions. As the number of regions increases, the MRxMHD solution appears to approach the ideal MHD solution assuming a continuum of nested flux surfaces. We demonstrate agreement between the numerical solutions obtained from the two codes through a convergence study.
\end{abstract}
\maketitle

\section{Introduction}
Ideal magnetohydrodynamics (MHD) permits solutions with singular current sheets.\citep{Parker1994} General three-dimensional (3D) ideal MHD
equilibria with a continuum of nested flux surfaces, as often assumed by stellarator
equilibrium solvers such as VMEC\citep{HirshmanW1983} and NSTAB,\citep{Garabedian2002} are susceptible
to the formation of singular current sheets at rational surfaces.\citep{Grad1967,BhattacharjeeHHNS1995,Helander2014} 
Nominally two-dimensional (2D) systems such as tokamaks can also develop
singular current sheets when subjected to resonant magnetic perturbations
(RMPs). The formation of ideal MHD singular current sheets has significant
practical implications. With a finite resistivity or other non-ideal
effects that enable magnetic reconnection, magnetic field lines surrounding
the ideal singular current sheets will break and reconnect, thereby releasing
magnetic energy; consequently, the magnetic field will evolve into a field with magnetic islands and possibly regions of stochastic
field lines if islands overlap.\citep{Biskamp1993,Biskamp2000} The sites of ideal MHD singular current sheets, therefore, serve as an indicator of where magnetic reconnection will occur. The intensities of the current sheets also measure the amount of energy available for reconnection.

\begin{figure}
\begin{centering}
\includegraphics[width=1\columnwidth]{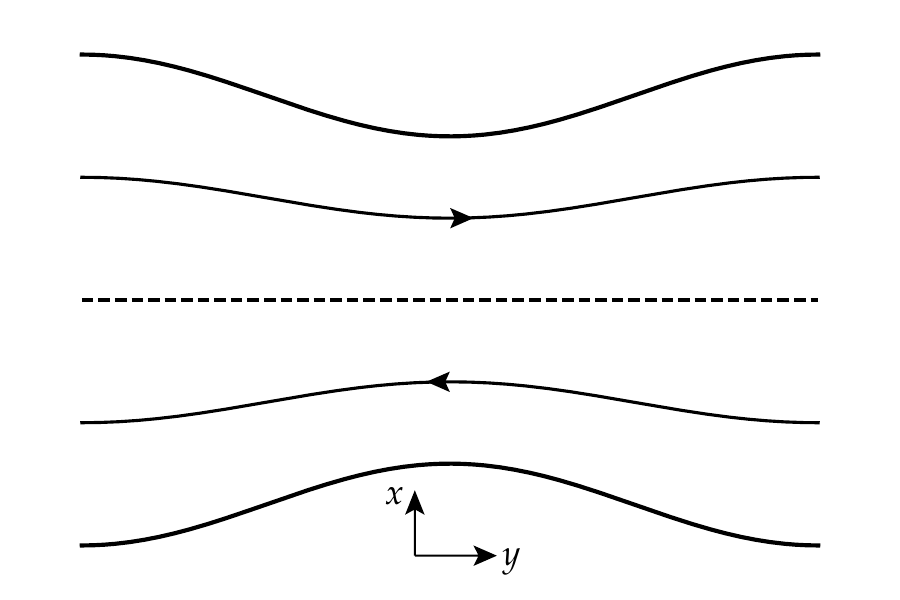}
\par\end{centering}
\caption{A sketch of the Hahm-Kulsrud-Taylor problem. The in-plane components
of the magnetic field reverse directions at the mid-plane (the dashed
line). The upper and lower boundaries are shaped by mirror-symmetric
sinusoidal perturbations. In response to the perturbation, a singular
current sheet develops at the mid-plane. \label{fig:A-sketch-of-HKT}}
\end{figure}

A prototype for singular current sheet formation driven by RMPs is
the Hahm-Kulsrud-Taylor (HKT) problem, \citep{HahmK1985,WangB1992a,DewarHBY2017} shown in
Figure \ref{fig:A-sketch-of-HKT}. This 2D problem has a magnetized
plasma enclosed by two conducting walls in slab geometry. Before
the conducting walls are perturbed, the initial magnetic field is
a smooth function of space. The in-plane component points along the $y$
direction and $B_{y}$ reverses direction at the mid-plane (the dashed
line in Figure \ref{fig:A-sketch-of-HKT}). A non-uniform $B_{z}$
component renders the magnetic field force-free. We then impose a
sinusoidal perturbation with an up-down symmetry to the conducting
walls and look for a new ideal equilibrium that is consistent with the boundary perturbation while conserving magnetic fluxes between flux surfaces. In this new equilibrium, a singular current sheet will develop at the mid-plane, which is a flux surface that resonates with the boundary perturbation. 

The primary objectives of this work are  (1) investigating the nature of
ideal singular current sheets by constructing numerical solutions of the HKT problem as an example and (2) assessing the accuracy of the numerical solutions via convergence tests. We limit ourselves to the case of vanishing plasma pressure in this paper, which results in a Dirac $\delta$-function current singularity. For
more general cases with a non-vanishing pressure gradient, a Pfirsch--Schl\"uter
current density that diverges algebraically towards the resonant surfaces
could arise in addition to the $\delta$-function current singularities.\citep{Helander2014}
We leave the Pfirsch--Schl\"uter current singularity to a future
study. 

This study employs two numerical codes: (1) a flux conserving Grad-Shafranov
(GS) solver \citep{HuangBZ2009} and (2) the Stepped Pressure Equilibrium
Code (SPEC).\citep{HudsonDDHMNL2012} 

The GS solver assumes a continuum of nested flux surfaces, which precludes magnetic island formation. By prescribing the toroidal (i.e., out-of-plane) and poloidal (i.e., in-plane) fluxes, the geometry of flux surfaces determines the magnetic field. The geometry of the flux surfaces is described by a mapping from coordinate space to physical space. The numerical implementation discretizes the mapping with a Chebyshev-Fourier pseudospectral method,\citep{Fornberg1995a,Trefethen2000} where the residual MHD force $\boldsymbol{J}\times\boldsymbol{B} - \nabla p$ is calculated on a set of collocation points. Here, we use standard notations for the magnetic field ($\boldsymbol{B}$), the electric current density ($\boldsymbol{J}$), and the plasma pressure ($p$). The collocation points are uniformly spaced along the Fourier ($y$) direction and correspond to the interior Chebyshev--Lobato points along the $x$ direction. The mapping is iteratively updated by an energy descent algorithm, similar to that of VMEC, until the residual MHD force is below a threshold.  

Previously, numerical solutions of the ideal HKT problem from the GS solver have been tested, showing agreement with the solutions of a fully Lagrangian solver\citep{ZhouHQB2016} and analytic solutions obtained with an asymptotic boundary-layer analysis.\citep{ZhouHRQB2019} However, the accuracy of the GS solution has not been fully assessed and quantified. The Lagrangian solver does not yield a converged magnetic field at the resonant surface and therefore cannot facilitate quantification of errors.  The boundary-layer analytic solution also cannot be used to assess the accuracy of the GS solution, because it is approximate and not exact.

To further assess the accuracy of the GS solution, it is not sufficient to rely on self-convergence. Even if the GS solution converges as the resolution increases, there is no guarantee that the converged solution is correct. To address this issue, we employ SPEC as an independent solver to benchmark the GS solver. Another motivation for employing SPEC in this study is that SPEC can handle a much broader class of 3D configurations. If SPEC can obtain approximate solutions to the ideal HKT problem, it can potentially be applied to more complicated 3D problems involving multiple resonant surfaces.

The SPEC code solves for multi-region
relaxed magnetohydrodynamic (MRxMHD) equilibria.\citep{HudsonDDHMNL2012}
MRxMHD does not assume a continuum of nested flux surfaces. Instead, the physical domain is divided into nested regions. In each region, the magnetic field relaxes to a Taylor state,\citep{Taylor1974} i.e., a Beltrami field satisfying the condition $\nabla\times\boldsymbol{B}=\mu\boldsymbol{B}$, where $\mu$ is a constant, while conserving magnetic helicity as well as the poloidal and the toroidal magnetic fluxes. Force-balance conditions are enforced across the interfaces between adjacent regions. Within each MRxMHD region, formation of magnetic islands and stochastic field line regions is allowed;\footnote{However, note that stochastic field line regions are not possible for the 2D HKT problem even when magnetic reconnection is allowed; only magnetic islands are possible.} and the interfaces between MRxMHD regions serve as ideal flux surfaces that prevent the magnetic field from relaxing to a global Taylor state. MRxMHD can be viewed as a bridge between Taylor's relaxation theory and ideal MHD. When there is only one region in the entire domain, MRxMHD is equivalent to Taylor's relaxation. On the other hand, in the limit of an infinite number of regions such that the ideal interfaces become a continuum, it has been shown that MRxMHD approaches ideal MHD under some conditions.\citep{DennisHDH2013, QuHDLH2021}

If these conditions hold, we expect SPEC solutions to approach the ideal MHD solution as the number of regions increases. Hence, we should be able to use SPEC solutions with a large number of regions to benchmark the GS solutions. However, Loizu \emph{et al.}\citep{LoizuHBLH2015} previously studied a similar problem of imposing an $m=2$, $n=1$
perturbation on a cylindrical screw pinch with SPEC and concluded
that a minimal finite jump, approximately proportional to the perturbation
amplitude, in the rotational transform across the resonant surface
is a \emph{sine qua non} condition for the existence of a solution.
Because the HKT problem has a continuous rotational transform, the
\emph{sine qua non} condition raises the question of whether the solutions
previously obtained with the fully Lagrangian solver and the GS solver
can also be obtained by SPEC. As it turns out, in this study we find that SPEC can actually obtain solutions to the HKT problem without requiring a discontinuous rotational transform; therefore, the previous interpretation of the \emph{sine qua non} condition as a necessary condition for the existence of a solution is not valid.

This paper is organized as follows. In Sec.~\ref{sec:Grad-Shafranov-solutions},
we briefly describe the flux preserving formulation of the GS
equation and review the linear and nonlinear solutions of the HKT problem. In Sec.~\ref{GS-solution}, we present numerical solutions and convergence tests from the GS solver. In Sec.~\ref{sec:SPEC-solutions}, the numerical
solutions and convergence tests from SPEC are presented. In Sec.~\ref{sec:Discussion}, we further examine the nature of the singular solution and discuss possible reasons for why SPEC failed to find solutions in Ref.~[\onlinecite{LoizuHBLH2015}] when the \emph{sine qua non} condition was not satisfied, as well as the correct interpretation of the \emph{sine qua non} condition. We present a case when the \emph{sine qua non} condition is marginally satisfied to demonstrate how that affects the nature of the solution. Finally, we conclude
and discuss future perspectives in Sec.~\ref{sec:Conclusions}.

\section{Grad--Shafranov Formulation of the Hahm-Kulsrud-Taylor Problem\label{sec:Grad-Shafranov-solutions}}

Two-dimensional MHD equilibria in  Cartesian geometry satisfy the
Grad-Shafranov equation 
\begin{equation}
\nabla^{2}\psi=-\frac{dP}{d\psi},\label{eq:GS}
\end{equation}
where 
\begin{equation}
P=p+\frac{B_{z}^{2}}{2}\label{eq:pressure}
\end{equation}
is a function of $\psi$. Here, the Cartesian coordinate $z$ is the
direction of translational symmetry. The flux function $\psi$ determines
the perpendicular components of the magnetic field through the relation
\begin{equation}
\boldsymbol{B}_{\perp}=\boldsymbol{\hat{z}}\times\nabla\psi.\label{eq:B_perp}
\end{equation}
Both the out-of-plane component $B_{z}$ and the plasma pressure $p$
are functions of $\psi$. The component $B_{z}$ is determined by
the conservation of magnetic flux. In this study, we set $p$ equal to zero.

With the magnetic fluxes prescribed, the magnetic field is determined
by the geometry of the flux surfaces. We can label the flux surfaces
with an arbitrary variable, and a convenient choice is to use the initial
positions $x_{0}$ of flux surfaces before the boundary perturbation is imposed.
The flux surfaces are described by a mapping from $\left(x_{0},y\right)$
to $\left(x,y\right)$ via a function $x\left(x_{0},y\right)$. Using the chain rule, we can express the partial derivatives with respect to the Cartesian coordinates in terms of the partial derivatives with respect to
the coordinates $\left(x_{0},y\right)$:
\begin{equation}
\left(\frac{\partial}{\partial x}\right)_{y}=\dfrac{1}{\partial x/\partial x_{0}}\frac{\partial}{\partial x_{0}},\label{eq:partial_x}
\end{equation}
\begin{equation}
\left(\frac{\partial}{\partial y}\right)_{x}=\frac{\partial}{\partial y}-\dfrac{\partial x/\partial y}{\partial x/\partial x_{0}}\frac{\partial}{\partial x_{0}}.\label{eq:partial_y}
\end{equation}
Here, the subscripts of the partial derivatives on the left-hand side indicate the coordinates that are held fixed; the partial derivatives on the right-hand side are with respect to the $\left(x_{0},y\right)$ coordinates. Hereafter, partial derivatives are taken to be with respect to the $\left(x_{0},y\right)$ coordinates by default, unless otherwise indicated by the subscripts. 

Using these relations, the Cartesian components of the in-plane magnetic field are given by
\begin{equation}
B_{x}=-\left(\frac{\partial\psi}{\partial y}\right)_{x}=\dfrac{\partial x/\partial y}{\partial x/\partial x_{0}}\frac{d\psi}{dx_{0}}\label{eq:Bx}
\end{equation}
and 
\begin{equation}
B_{y}=\left(\frac{\partial\psi}{\partial x}\right)_{y}=\dfrac{1}{\partial x/\partial x_{0}}\frac{d\psi}{dx_{0}}.\label{eq:By}
\end{equation}
The out-of-plane component $B_{z}$ is determined by conservation
of magnetic flux as 
\begin{equation}
B_{z}\left(x_{0}\right)=\frac{B_{z0}\left(x_{0}\right)}{\left\langle \dfrac{\partial x}{\partial x_{0}}\right\rangle }.\label{eq:Bz}
\end{equation}
Here, $B_{z0}$ is the initial $z$-component of the magnetic field;
the flux surface average $\left\langle f\right\rangle $ is defined
as 
\begin{equation}
\left\langle f\right\rangle \equiv\frac{1}{L}\int_{0}^{L}f\left(x_{0},y\right)dy\label{eq:ave}
\end{equation}
for an arbitrary function $f\left(x_{0},y\right)$, with $y\in$$\left[0,L\right]$
being the domain of the system along the $y$ direction. The out-of-plane
component of the current density is given by
\begin{equation}
J_{z}=\nabla^{2}\psi=\left(\frac{d\psi}{dx_{0}}\right)^{-1}\frac{\partial}{\partial x_{0}}\left(\frac{B_{x}^{2}+B_{y}^{2}}{2}\right)-\frac{\partial B_{x}}{\partial y},\label{eq:Jz}
\end{equation}
and the GS equation can be written as 
\begin{equation}
H=-\left(\frac{d\psi}{dx_{0}}\right)^{-1}\frac{\partial}{\partial x_{0}}\left(\frac{B_{x}^{2}+B_{y}^{2}}{2}+P\right)+\frac{\partial B_{x}}{\partial y}=0.\label{eq:GS1}
\end{equation}
The residual MHD force is given by
\begin{equation}
\boldsymbol{F}\equiv H\nabla\psi.\label{eq:force}
\end{equation}
To obtain the solution, we can use $F_{x}=HB_{y}$ to push the flux surfaces along the $x$ direction, subjected to a friction force to damp the energy until the system settles down to an equilibrium. 

For the HKT problem, we consider an initial force-free equilibrium
\begin{equation}
\boldsymbol{B}_{0}=x_0\boldsymbol{\hat{y}}+\sqrt{B_{0}^{2}-x_0^{2}}\boldsymbol{\hat{z}}\label{eq:initial_B}
\end{equation}
in the domain $x_0\in\left[-a,a\right]$ and $y\in\left[0,L\right]$,
where the $y$ direction is assumed to be periodic. The corresponding in-plane flux function is 
$\psi=x_{0}^{2}/2$. We impose a sinusoidal perturbation on the boundary that deforms $x=\pm a$ to $x=\pm\left(a+\delta\cos\left(ky\right)\right)$
and let the system evolve under the constraints of ideal MHD to a new equilibrium.

For a small boundary perturbation, we may linearize the GS equation
in terms of the displacements of the flux surfaces $\xi\left(x_{0},y\right)\equiv x\left(x_{0},y\right)-x_{0}$.
To the leading order in $\xi$, the magnetic field components are
\begin{equation}
B_{x}\simeq\frac{\partial\xi}{\partial y}\frac{d\psi}{dx_{0}},\label{eq:Bx1}
\end{equation}
\begin{equation}
B_{y}\simeq\left(1-\partial\xi/\partial x_{0}\right)\frac{d\psi}{dx_{0}},\label{eq:By1}
\end{equation}
and 
\begin{equation}
B_{z}\simeq B_{z0}\left(1-\left\langle \frac{\partial\xi}{\partial x_{0}}\right\rangle \right).\label{eq:Bz1}
\end{equation}
The linearized GS equation now reads
\begin{equation}
\frac{\partial}{\partial x_{0}}\left(\left(\frac{d\psi}{dx_{0}}\right)^{2}\frac{\partial\xi}{\partial x_{0}}+B_{z0}^{2}\left\langle \frac{\partial\xi}{\partial x_{0}}\right\rangle \right)+\frac{\partial^{2}\xi}{\partial y^{2}}\left(\frac{d\psi}{dx_{0}}\right)^{2}=0.\label{eq:GS-linear}
\end{equation}

For the HKT problem with $\psi=x_{0}^{2}/2$ and the boundary condition
$\xi\left(\pm a,y\right)=\pm\delta\cos\left(ky\right)$, if we adopt
the ansatz $\xi=\bar{\xi}(x_{0})\cos(ky)$, then $\left\langle \partial\xi/\partial x_{0}\right\rangle =0$
and the linearized GS equation reduces to
\begin{equation}
\frac{d^{2}}{dx_{0}^{2}}\left(x_{0}\bar{\xi}\right)-k^{2}x_{0}\bar{\xi}=0.\label{eq:GS-linear_HKT}
\end{equation}
The general solution of Eq.~(\ref{eq:GS-linear_HKT}) is a linear superposition of two independent
solutions
\begin{equation}
\bar{\xi}=c_{1}\frac{\sinh\left(k\left|x_{0}\right|\right)}{x_{0}}+c_{2}\frac{\cosh\left(kx_{0}\right)}{x_{0}},\label{eq:linear_sol}
\end{equation}
and the boundary condition $\bar{\xi}(\pm a)=\pm\delta$ requires
\begin{equation}
\delta=\frac{c_{1}\sinh\left(ka\right)+c_{2}\cosh\left(ka\right)}{a}.\label{eq:BC-2}
\end{equation}

We can immediately see that the linear solution is problematic near
the resonant surface at $x_{0}=0$. The divergence of $\cosh\left(kx_{0}\right)/x_{0}$
at $x_{0}=0$ suggests that the coefficient $c_{2}$ must be set to
zero, and the boundary condition (\ref{eq:BC-2}) then determines
the coefficient $c_{1}=a\delta/\sinh\left(ka\right)$. However, the
limit that $\lim_{x_{0}\to0}\sinh\left(k\left|x_{0}\right|\right)/x_{0}=k$
yields $x\simeq x_{0}+\left(ka\delta/\sinh\left(ka\right)\right)\cos\left(ky\right)$
in the vicinity of $x_{0}=0$, leading to overlap of flux
surfaces when $\left|x_{0}\right|\leq ka\delta/\sinh\left(ka\right)$, which amounts to a physical inconsistency and is unpermitted.
Therefore, within an inner region $\left|x_{0}\right|\lesssim\mathcal{O}\left(ka\delta/\sinh\left(ka\right)\right)$,
the linear solution is not valid and we must consider the nonlinear
solution.

The nonlinear solution of the inner region was first derived by Rosenbluth,
Dagazian, and Rutherford (hereafter RDR) for the ideal internal kink
instability \citep{RosenbluthDR1973} and was later adapted to the HKT
problem.\citep{BoozerP2010,ZhouHRQB2019} Because $d\psi/dx_{0}\to0$
in the inner region, the dominant balance of the GS equation (\ref{eq:GS1})
is approximately given by 

\begin{equation}
\frac{\partial}{\partial x_{0}}\left(\frac{B_{y}^{2}}{2}+P\left(x_{0}\right)\right)=0;\label{eq:GS_inner}
\end{equation}
here, we have neglected $B_{x}^{2}$ compared to $B_{y}^{2}$ in
Eq. (\ref{eq:GS1}) by assuming $\left|\partial x/\partial y\right|\ll1$.
Integrating Eq.~(\ref{eq:GS_inner}) yields 
\begin{equation}
B_{y}=\dfrac{1}{\partial x/\partial x_{0}}\frac{d\psi}{dx_{0}}=\sign\left(\frac{d\psi}{dx_{0}}\right)\sqrt{f(x_{0})+g(y)},\label{eq:Inner_sol}
\end{equation}
where 
\begin{equation}
f\left(x_{0}\right)=-2P\left(x_{0}\right)+\text{const}\label{eq:constrant}
\end{equation}
and $g(y)$ is an arbitrary function that will be determined later
by asymptotic matching to the outer solution; the $\sign\left(d\psi/dx_{0}\right)$
factor comes from the requirement that $\partial x/\partial x_{0}>0$
must be satisfied to avoid overlapping flux surfaces. Without loss of generality, we are free to set $f(0)=0$, and the tangential discontinuity of $B_{y}$ at $x_{0}=0$ is 
\begin{equation}
\left.B_{y}\right|_{0^{\pm}}=\pm\sqrt{g(y)}.\label{eq:tangential_discontinuity}
\end{equation}
Using the flux function $\psi=x_{0}^{2}/2$ for the HKT problem and
integrating Eq. (\ref{eq:Inner_sol}) one more time yields the inner
solution of RDR
\begin{equation}
x_{RDR}\left(x_{0},y\right)=\int_{0}^{x_{0}}\frac{\left|x'\right|}{\sqrt{f(x')+g(y)}}dx'.\label{eq:RDR}
\end{equation}
Note that the functions $f\left(x_{0}\right)$ and $g\left(y\right)$
are not independent, but are related through Eq.~(\ref{eq:constrant}).
Here, $P=B_{z}^{2}/2$ and $B_{z}$ is determined by Eq.~(\ref{eq:Bz}) with $x\left(x_{0}, y\right)$ replaced by $x_{RDR}\left(x_{0}, y\right)$.
The resulting relation is cumbersome. To simplify the problem, we
further assume that $\left|B_{z}\right|\gg\left|B_{\perp}\right|$ (i.e., in the so-called reduced MHD regime) and replace the constraint (\ref{eq:constrant}) by the incompressible constraint $\left\langle \partial x/\partial x_{0}\right\rangle =1$, yielding
\begin{equation}
\left\langle \frac{1}{\sqrt{f(x_{0})+g(y)}}\right\rangle =\frac{1}{|x_{0}|}.\label{eq:incompressible}
\end{equation}
Once $g(y)$ is obtained, the constraint (\ref{eq:incompressible}) then determines $f\left(x_{0}\right)$.

The function $g(y)$ can be obtained via asymptotic matching to the
linear solution in the outer region. The readers are referred to Ref.~[\onlinecite{ZhouHRQB2019}] for further detail of adapting the matching
method of RDR for the internal kink mode to the HKT problem. Here,
we simply quote the relevant results. The function $g(y)$ can be
obtained by numerically solving an integral equation,\citep{LoizuH2017}
but a good analytic approximation for $g(y)$ is \textbf{
\begin{equation}
g(y)\simeq\frac{4c_{1}^{2}k^{2}}{3}\sin^{8}(ky/2),\label{eq:gy}
\end{equation}
}where $c_{1}$ is the coefficient of the outer solution given
by
\begin{equation}
c_{1}=\frac{12\sinh(ka)-\sqrt{144\sinh^{2}(ka)-168 a \delta k^{2}\cosh(ka)}}{7k^{2}\cosh(ka)}.\label{eq:c1}
\end{equation}

\section{Numerical Solutions of the Grad-Shafranov Equation} \label{GS-solution}

\begin{figure}
\begin{centering}
\includegraphics[width=1\columnwidth]{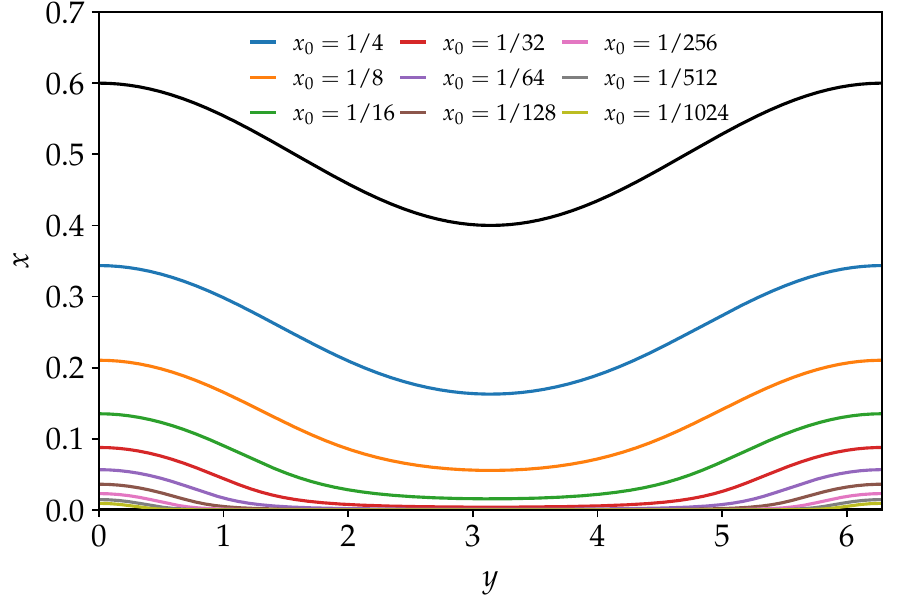}
\par\end{centering}
\caption{A selection of flux surfaces obtained from the highest resolution
calculation that we use as the reference for convergence tests in
this study.\label{fig:flux-surfaces}}
\end{figure}

Now we present numerical solutions to the HKT problem obtained by the GS solver. We set the free parameters of this problem to $B_{0}=10$, $a=1/2$, $\delta=0.1$, and $k=2\pi/L=1$. We assume a mirror symmetry of the solution and solve in only half of the domain $x_0\in[0,1/2]$.

We perform two sets of numerical calculations. The first set employs a direct Chebyshev-Fourier pseudospectral discretization of the GS equation. For the second set, we take advantage of the knowledge of RDR's analytic solution and express the geometry of flux surfaces as $x(x_{0},y)=x_{\text{RDR}}(x_{0},y)+\tilde{x}(x_{0},y)$. Here, to calculate the analytic solution $x_{RDR}$, we adopt the analytic approximation (\ref{eq:gy}) for $g(y)$ and numerically solve the incompressible constraint (\ref{eq:incompressible}) to obtain $f\left(x_{0}\right)$. We then
numerically integrate Eq.~(\ref{eq:RDR}) to obtain $x_{RDR}$.
We rewrite the GS equation in terms of the deviation $\tilde{x}$ from the RDR solution and implement a special version of the GS solver for this formulation. Because the analytic solution accounts for most of the singular behavior near the resonant surface, the accuracy of the second set of solutions is substantially improved.

We use $N_{y}=512$ collocation points along the $y$ direction to
ensure that most of the numerical errors are due to the discretization
along the $x_{0}$ direction. We then test the convergence of the
numerical solution by increasing the number of Chebyshev collocation
points $N_{x}$. We perform calculations with $N_{x}=8$, 16, 32,
64, and 128. The Chebyshev collocation points cluster near the edges
of the domain, with the shortest distance between  the  collocation points scales as $1/N_{x}^{2}$.
For $N_{x}=128$, the closest collocation point is at $x_{0}=7.5\times10^{-5}$.
Due to the lack of a perfectly precise solution for the convergence
test, we take the most accurate numerical solution available
as a substitute. For that purpose, the $N_{x}=128$ solution from
the second set (with the subtraction of the RDR solution) serves as
the reference.

Our primary diagnostics for the convergence test are: (a) the discontinuity
of magnetic field at the resonant surface $\left.B_{y}\right|_{0^{+}}$
($\left.B_{y}\right|_{0^{-}}=-\left.B_{y}\right|_{0^{+}}$ from symmetry); and
(b) the geometry of a selection of flux surfaces. For the latter,
we use the flux surfaces labeled by $x_{0}=1/4$, $1/8$, $1/16$,$\ldots$, $1/1024$. This set of flux surfaces is shown in Fig.~\ref{fig:flux-surfaces}. We quantify the errors of a solution by the $L_{2}$ norms of the
differences of relevant quantities relative to the reference solution.
Specifically, we use 
\begin{equation}
\left\Vert \left.\Delta B_{y}\right|_{0^{+}}\right\Vert _{2}\equiv\left\langle \left(\left.B_{y}\right|_{0^{+}}-\left.B_{y}^{ref}\right|_{0^{+}}\right){}^{2}\right\rangle ^{1/2}\label{eq:By_err}
\end{equation}
and 
\begin{equation}
\left\Vert \Delta x\right\Vert _{2}\equiv\left\langle \left(x-x^{ref}\right){}^{2}\right\rangle ^{1/2},\label{eq:x_err}
\end{equation}
where the flux surface average is defined in Eq.~(\ref{eq:ave}).

The calculation of $B_{y}$ using Eq.~(\ref{eq:By}) fails at the resonant
surface, because both the denominator and the numerator approach zero.
To obtain $\left.B_{y}\right|_{0^{+}}$, we perform a polynomial extrapolation
using the barycentric formula\citep{BerrutT2004} with values of $B_{y}$
on all the collocation points other than $x_{0}=0$. Additionally, because
the flux surfaces of choice for the convergence test do not coincide
with the Chebyshev collocation points, we have to perform a polynomial
interpolation to determine their geometry.  

\begin{figure}
\begin{centering}
\includegraphics[width=1\columnwidth]{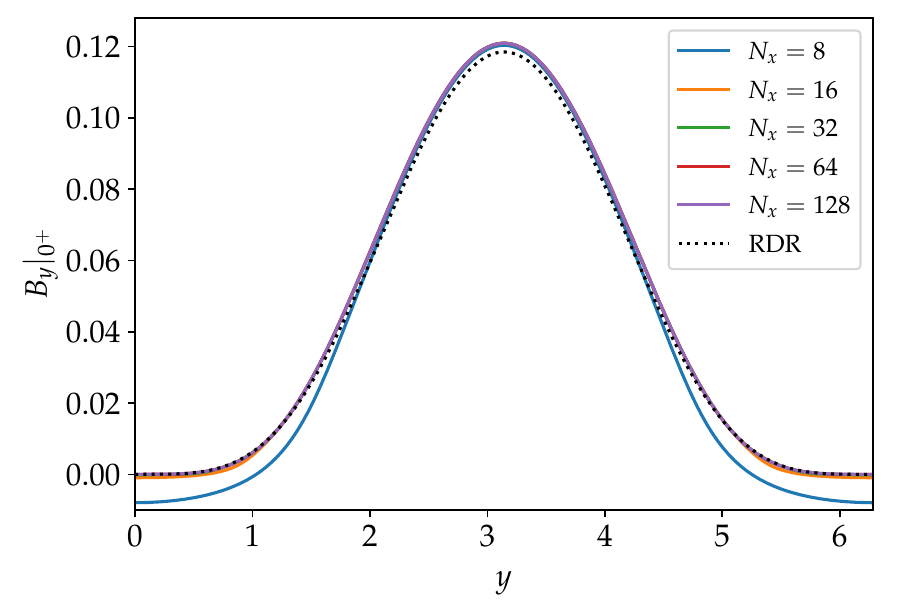}
\par\end{centering}
\caption{The magnetic field $B_{y}|_{0^{+}}$ at the lower boundary of the
computational domain obtained from the GS solver without RDR subtraction. For reference, the dotted line shows the RDR solution.
\label{fig:The-magnetic-field}}
\end{figure}

Figure \ref{fig:The-magnetic-field} shows $\left.B_{y}\right|_{0^{+}}$
from the GS solver with increasing grid resolutions. For $N_{x}=8$,
we can see that $\left.B_{y}\right|_{0^{+}}$ becomes negative near
$y=0$ and $y=2\pi$. This is a numerical error due to discretization and extrapolation, as the true solution should remain positive and only becomes zero at $y=0$ and $y=2\pi$. As $N_{x}$ increases, the solution quickly converges and the curves are virtually on top of one another when $N_{x}\ge16$. The values near $y=0$ and $y=2\pi$ remain slightly
negative, but the magnitude rapidly decreases as $N_{x}$ increases.
For the second set of solutions with RDR subtraction, the curves virtually
overlap with each other for all the cases we have done (not shown). The dotted line in Fig.~\ref{fig:The-magnetic-field} shows the RDR solution. We can see that although the RDR solution is a good approximation, there is a visible difference between the RDR solution and the converged GS solution.

\begin{figure}
\begin{centering}
\includegraphics[width=1\columnwidth]{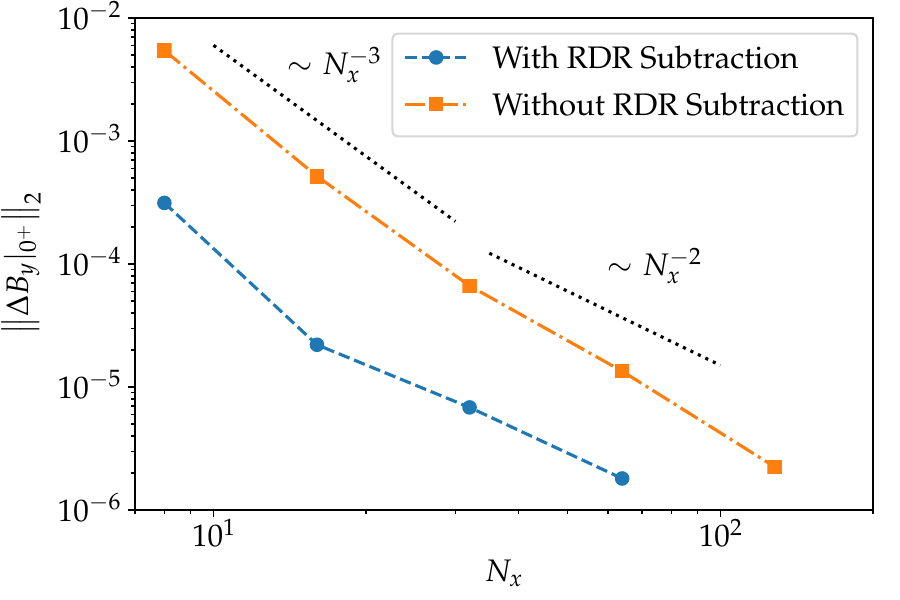}
\par\end{centering}
\caption{Convergence of $\left.B_{y}\right|_{0^{+}}$ errors from the GS solvers
with and without subtracting the RDR solution. \label{fig:Convergence-of-By-err}}

\end{figure}

\begin{figure}
\begin{centering}
\includegraphics[width=1\columnwidth]{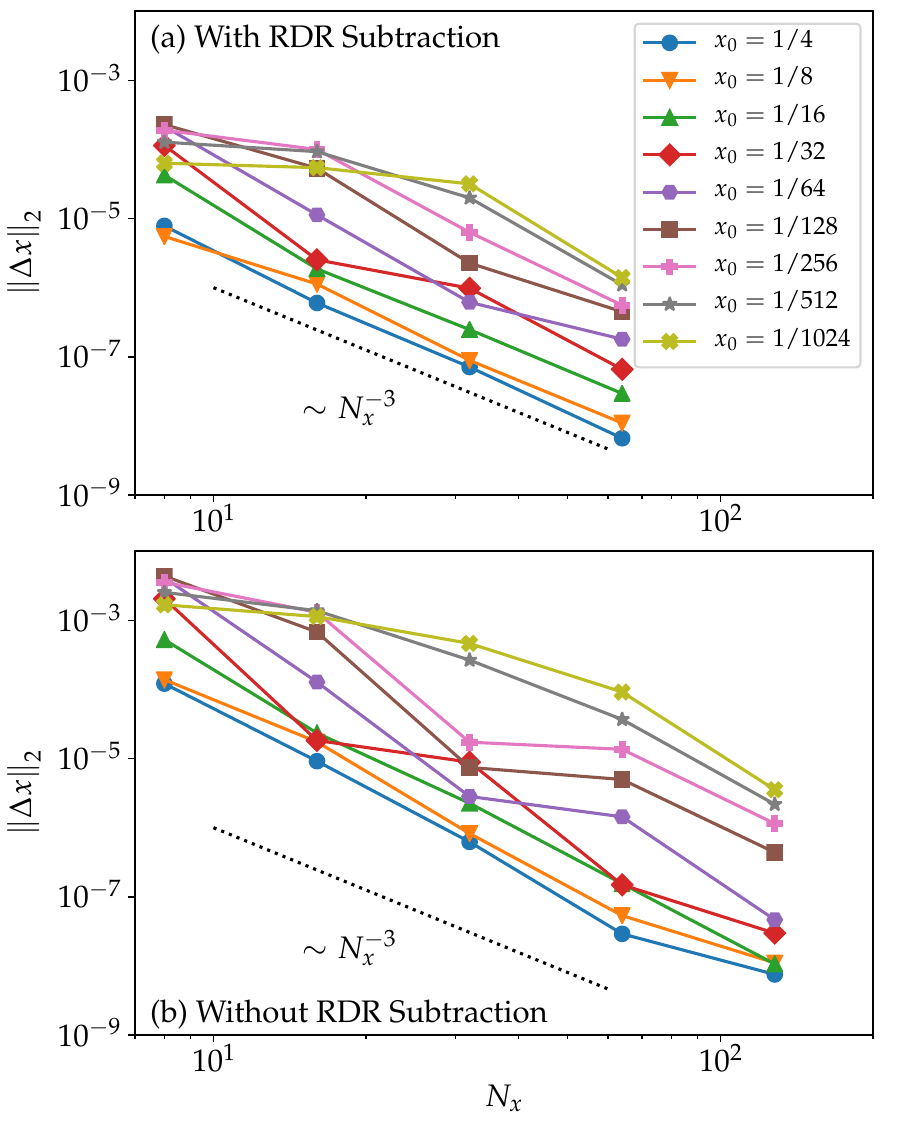}
\par\end{centering}
\caption{Convergence of flux surface errors from the GS solvers with
and without subtracting the RDR solution. \label{fig:Convergence-of-flux-surface}}
\end{figure}

Figure \ref{fig:Convergence-of-By-err} shows the convergence of $\left.B_{y}\right|_{0^{+}}$
errors for both sets of solutions. We can see that applying RDR subtraction
reduces the errors by approximately one order of magnitude, but the
overall convergence rates are similar for both sets of solutions. Likewise, the convergence of flux surface errors is shown
in Figure \ref{fig:Convergence-of-flux-surface} for both sets of
solutions. Evidently, flux surfaces closer to the resonant surface
are more difficult to solve accurately. Again, the RDR subtraction reduces the errors by approximately an order of magnitude, but the
overall convergence rate remains similar. 

Note that the data points for $N_{x}=128$ with RDR subtraction are
missing in Figures \ref{fig:Convergence-of-By-err} and \ref{fig:Convergence-of-flux-surface},
because that solution serves as the reference. The convergence tests
provide a base for estimating the errors of the reference solution.
Because the same reference solution will also be used for the convergence
test of SPEC solutions, it is important to ensure that the reference solution is sufficiently accurate. By extrapolating the trends in Fig.~\ref{fig:Convergence-of-By-err} and Fig.~\ref{fig:Convergence-of-flux-surface}(a), we estimate the reference solution's error of $B_y|_{0^+}$ to be smaller than $10^{-6}$, error of the flux surface labeled by $x_0=1/4$ smaller than $10^{-9}$, and error of the flux surface labeled by $x_0=1/1024$ smaller than $10^{-6}$.

\section{SPEC solutions\label{sec:SPEC-solutions}}

Now we continue with the SPEC solutions to the HKT problem. Here we
also present the results from two sets of numerical calculations. For the first
set, the initial positions of interfaces between volumes are uniformly spaced before the boundary perturbation is imposed. We start from the number of volumes $N_{vol}=2$, then
increase to $N_{vol}=4$, $8$, up to $N_{vol}=128$. For the second set of calculations, we explore the possible advantages of packing more volumes near the resonant surface. Because the best strategy for packing volumes is not a priori clear, we adopt a procedure of refining only the nearest volume to the resonant surface to see how SPEC performs under this extreme scenario of local refinement. The procedure goes as follows: We start from $N_{vol}=2$. At each level of refinement, the volume adjacent to the resonant surface is divided into two equal volumes. In this way, we go up to an ``effective'' $N_{vol}=512$, meaning that the smallest volume is $1/512$ of the domain, while the actual number of volumes is $N_{vol}=10$. The interfaces between the volumes for the highest resolution case of the second set exactly correspond to the flux surfaces we use for convergence tests shown in Fig.~\ref{fig:flux-surfaces}.

We test the convergence of the two sets of SPEC solutions as the number of volumes increases, using the highest resolution GS solution as the reference. The number of Fourier harmonics along the $y$ direction is 48 for all the SPEC calculations presented here. 

When SPEC finds a solution, it is not guaranteed that the ideal interfaces between volumes will not overlap with each other. Overlapping ideal interfaces are not permitted on physical grounds, but they do occasionally occur in SPEC solutions, especially for those interfaces close to the resonant surface, and this will cause the SPEC algorithm to crash. Because SPEC uses Newton's method to find the solution, having a good initial guess is crucial. A useful approach to overcome the problem of overlapping ideal interfaces is to start from a small boundary perturbation, find the solution, then use the solution as the initial guess for a slightly increased boundary perturbation. This process is repeated until the full amplitude of boundary perturbation is reached. 

\begin{figure}
\begin{centering}
\includegraphics[width=1\columnwidth]{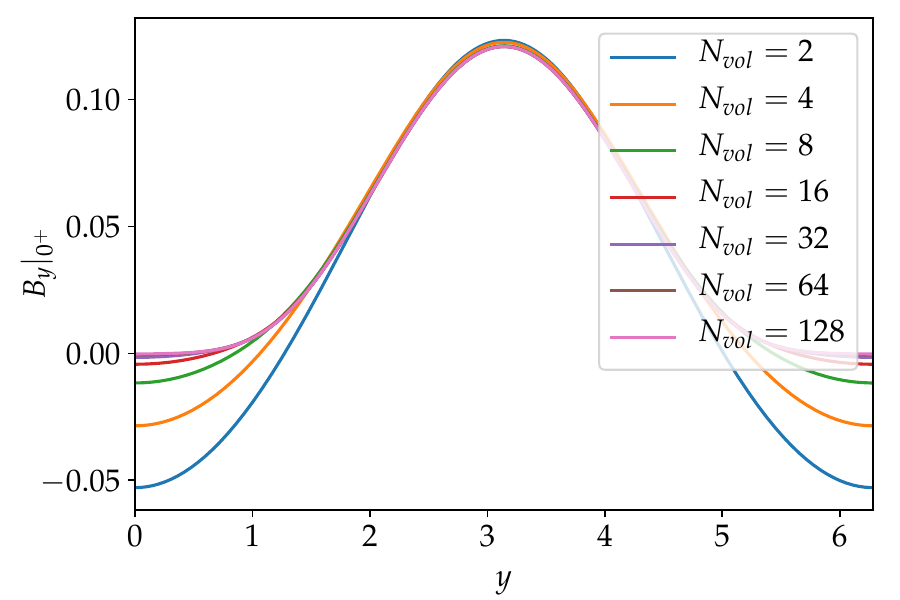}
\par\end{centering}
\caption{Convergence of $B_{y}|_{0^{+}}$ of SPEC solutions as the number of
uniformly spaced volumes increases.\label{fig:Convergence-of-BySPEC-uniform}}
\end{figure}

\begin{figure}
\includegraphics[width=1\columnwidth]{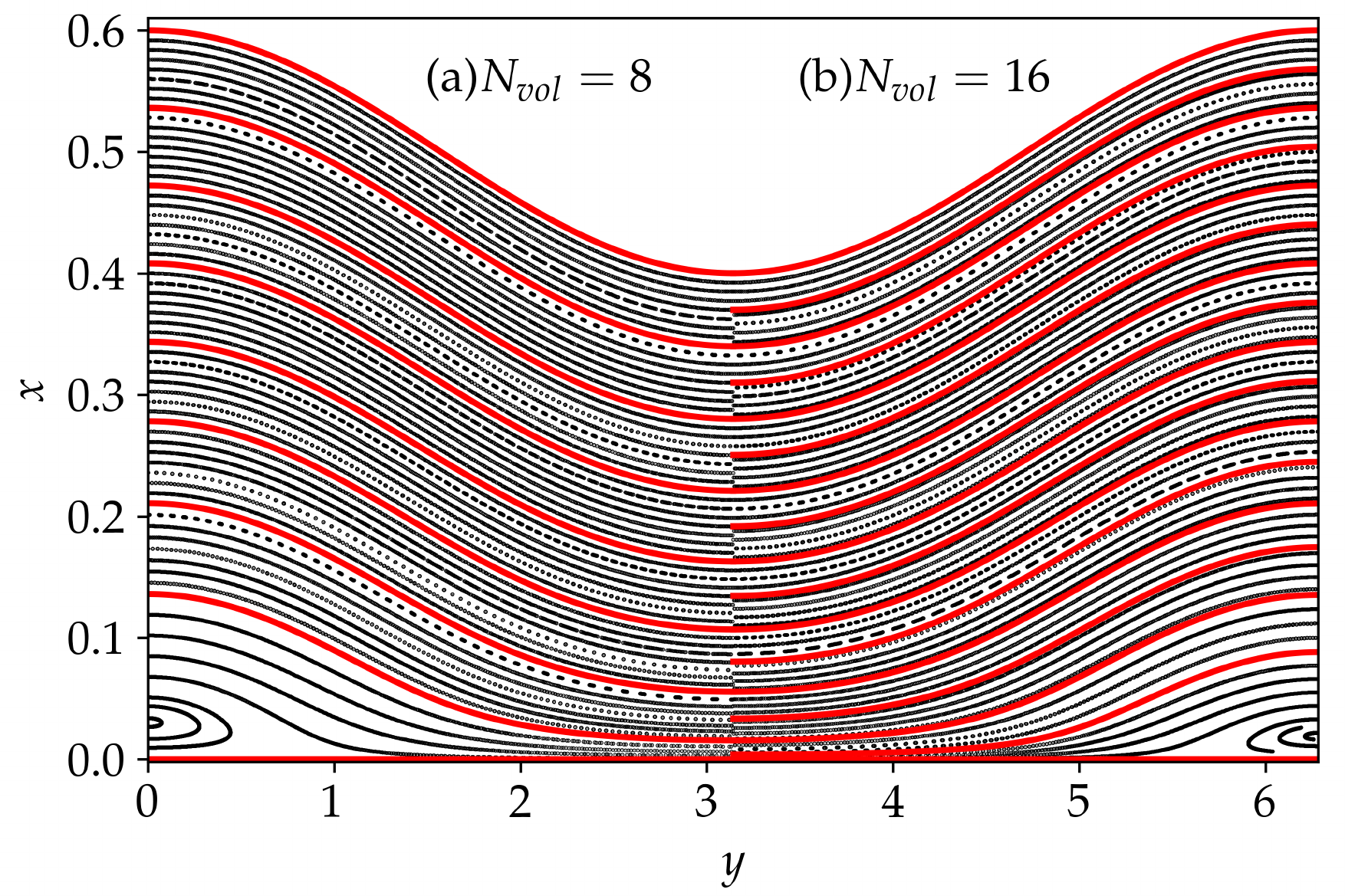}

\caption{Ideal interfaces between MRxMHD volumes (red) and samples of Poincar\'e 
plot in each volume (black). The left-hand side shows the case with
eight volumes, and the right-hand side shows the case with sixteen
volumes.\label{fig:KAM-interfaces-and-Poincare}}
\end{figure}

Figure \ref{fig:Convergence-of-BySPEC-uniform} shows $\left.B_{y}\right|_{0^{+}}$
from SPEC using uniformly-spaced volumes. Similar to the GS solutions
shown in Figure \ref{fig:The-magnetic-field}, the values of $\left.B_{y}\right|_{0^{+}}$
in SPEC solutions also become negative near $y=0$ and $y=2\pi$,
but the magnitude rapidly decreases as the number of volumes increases.
The reason for negative $\left.B_{y}\right|_{0^{+}}$ is the
presence of residual magnetic islands near the resonant surface,\citep{DewarHBY2017} as
we can see in Figure \ref{fig:KAM-interfaces-and-Poincare}. Here,
the red lines are the ideal interfaces and the black dots represent samples
of the Poincar\'e  plot from field line tracing. The left-hand-side
of the figure shows the $N_{vol}=8$ case, while the right-hand-side
shows the $N_{vol}=16$ case. The Poincar\'e  plot reveals the residual
islands in the lower left and the lower right corners. The size of
the island decreases as $N_{vol}$ increases from $8$ to $16$. This
trend continues as $N_{vol}$ further increases, resulting in the
decrease of the magnitude of negative $\left.B_{y}\right|_{0^{+}}$.

\begin{figure}
\begin{centering}
\includegraphics[width=1\columnwidth]{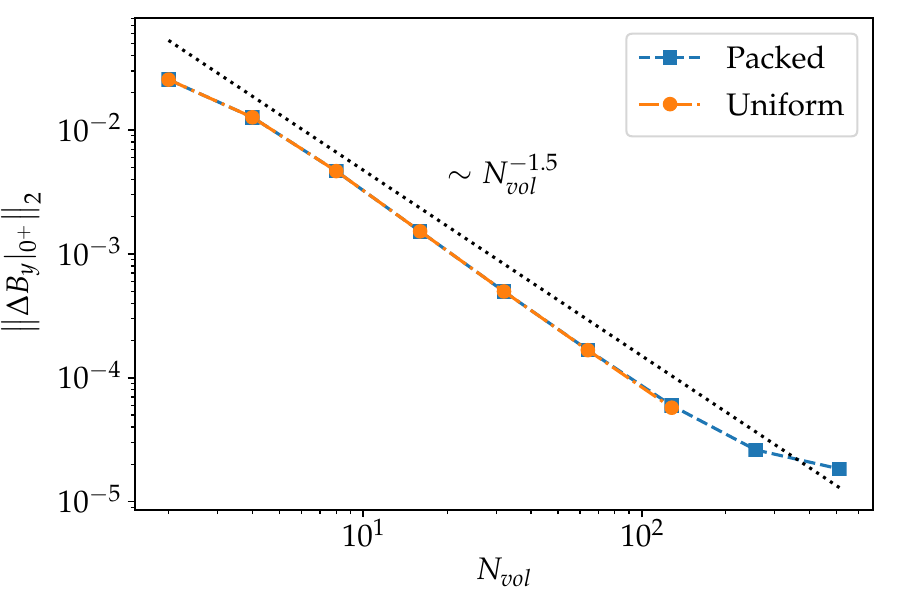}
\par\end{centering}
\caption{Convergence of the $\left.B_{y}\right|_{0^{+}}$ errors of SPEC solutions
for cases of packed volume and uniformly-spaced volumes.\label{fig:Convergence-of-the_By_error_SPEC}}
\end{figure}

\begin{figure}
\begin{centering}
\includegraphics[width=1\columnwidth]{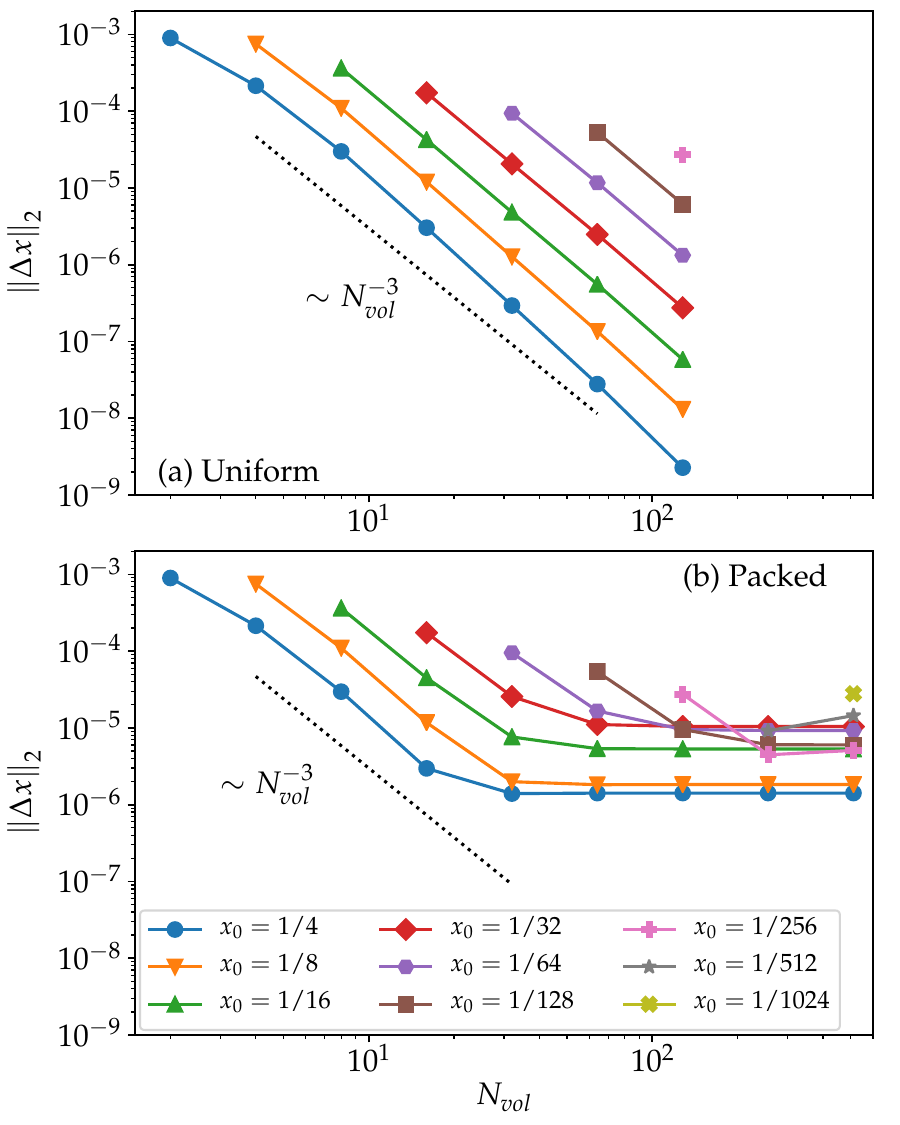}
\par\end{centering}
\caption{Convergence of the flux surface errors of SPEC solutions. Panel (a)
shows the cases of uniformly-spaced volumes, and panel (b) shows the
cases of packed volumes near the resonant surface. Note that $N_{vol}$ for the packed cases corresponds to the ``effective'' number of volumes as discussed in the text, not the actual number of volumes.  Because the outer region never gets refined for the packed cases, we do not expect the solutions to approach the ideal MHD solution even in the limit of $N_{vol}\to \infty$.}\label{fig:Convergence-of-the-surface-error_SPEC}
\end{figure}

Figure \ref{fig:Convergence-of-the_By_error_SPEC} shows the convergence
of the errors of $\left.B_{y}\right|_{0^{+}}$ as $N_{vol}$ increases,
for both sets of SPEC solutions. Here, for the cases of packed volumes,
$N_{vol}$ corresponds to the ``effective'' number of volumes as
discussed above. We can see that the errors from both sets are nearly
identical for the same $N_{vol}$, even though the volumes far away
from the resonant surface are much coarser for the packed cases. The
errors approximately scale as $N_{vol}^{-1.5}$ for both uniform
and packed cases. This finding suggests that $\left.B_{y}\right|_{0^{+}}$
may not strongly depend on the accuracy in the outer region. 

On the other hand, the effects of inadequate resolution in the outer
region are evident in the convergence of flux surface errors, shown
in Figure \ref{fig:Convergence-of-the-surface-error_SPEC}. Here,
the errors consistently scale as $N_{vol}^{-3}$ for cases of uniform
volumes. For cases of packed volumes, although the errors initially
decrease as $N_{vol}^{-3}$, the trend eventually flattens as $N_{vol}$
further increases.  It is possible that the stalling of convergence
in the outer region may eventually affect the convergence of $\left.B_{y}\right|_{0^{+}}$
for the packed cases. We can see that the last point of packed cases
in Figure \ref{fig:Convergence-of-the_By_error_SPEC} exhibits some deviation from the $N_{vol}^{-1.5}$ scaling. Another possible reason for the deviation is that SPEC solutions only use 48 Fourier modes, which may not be sufficient to accurately represent the flux surfaces near the resonant surface.

The results of packed-volume solutions show the effectiveness of local refinement, even when using the extreme refinement scenario adopted here. A better strategy in practical applications would be to refine over the entire domain while placing more volumes near the resonant surface. Nonetheless, our results indicate that a good approximation of the resonant singular current density may be obtained even with relatively coarse volumes away from the resonant surface.

\section{Discussion\label{sec:Discussion}}

\subsection{Nature of the singular solution}

The agreement between the solutions of the GS solver and SPEC suggests that both codes are approaching the true solution of the HKT problem as the resolution (or number of volumes) increases. Now we further examine the nature of the singular solution.

The finite tangential discontinuity $\left.B_{y}\right|_{0^{\pm}}$
arises from a continuous initial magnetic field through the compression
of the space between flux surfaces, which is evident from the flux
surfaces shown in Figure \ref{fig:flux-surfaces}. As we can infer
from the RDR solution (\ref{eq:RDR}), for flux surfaces sufficiently
close to the resonant surface such that the condition 
\begin{equation}
f\left(x_{0}\right)\ll g(y)\label{eq:condition}
\end{equation}
 is satisfied, we have 

\begin{equation}
x\left(x_{0},y\right)\simeq\frac{x_{0}^{2}}{\sqrt{g(y)}}.\label{eq:RDR-1}
\end{equation}
Because $f(0)=0$ and $g(y)\simeq\left(4c_{1}^{2}k^{2}/3\right)\sin^{8}(\pi y/L)$,
the condition (\ref{eq:condition}) will eventually be satisfied for
sufficiently small $x_{0}$ for all $y$ except at $y=0$ and $y=L$,
but the transition to the quadratic mapping $x\sim x_{0}^{2}$ occurs at different $x_{0}$ for different $y$. To compensate for the strong compression of the quadratic mapping, the ``downstream'' regions of flux surfaces near $y=0$ and $y=L$ have to bulge outward to maintain approximate incompressibility. 

Now we show that the flux surfaces sufficiently close to the resonant
surface satisfy a similarity relation near the downstream region
after a proper rescaling. To reveal the rescaling rules, we first
need to establish the behavior of $f\left(x_{0}\right)$ near $x_{0}=0$.
When the function $g(y)$ is known, the function $f(x_{0})$ can be
obtained by solving Eq.~(\ref{eq:incompressible}). Because $f(x_{0})\to0$
in the limit $|x_{0}|\to0$, the function $1/\sqrt{f(x_{0})+g(y)}$
is localized near $y=0$, $L.$ Hence, in this limit we can approximate
$g\left(y\right)$ by its leading order Taylor expansion, yielding
\begin{align}
 & \left\langle \frac{1}{\sqrt{f(x_{0})+g(y)}}\right\rangle \nonumber \\
\simeq & \frac{2}{L}\int_{0}^{\infty}\frac{dy}{\sqrt{f(x_{0})+\left(4c_{1}^{2}k^{2}/3\right)(ky/2)^{8}}}\nonumber \\
= & \frac{2}{\pi^{3/2}}\dfrac{\Gamma(3/8)\Gamma(9/8)}{f(x_{0})^{3/8}\left(4c_{1}^{2}k^{2}/3\right)^{1/8}},\label{eq:Ave}
\end{align}
where $\Gamma$ is the gamma function.\citep{AbramowitzS1972}
Plugging Eq.~(\ref{eq:Ave}) into Eq.~(\ref{eq:incompressible})
yields the leading order behavior of $f(x_{0})$ in the limit $\left|x_{0}\right|\to0$:
\begin{equation}
f(x_{0})\simeq\left[c_{f}\left|x_{0}\right|\right]^{8/3},\label{eq:f_0}
\end{equation}
where 
\begin{align}
c_{f} & \equiv\frac{2(3/4)^{1/8}}{\pi^{3/2}}\Gamma(3/8)\Gamma(9/8)\left(c_{1}k\right)^{-1/4}\nonumber \\
 & \simeq0.7735\left(c_{1}k\right)^{-1/4}.\label{eq:cf}
\end{align}

\begin{figure}
\begin{centering}
\includegraphics[width=1\columnwidth]{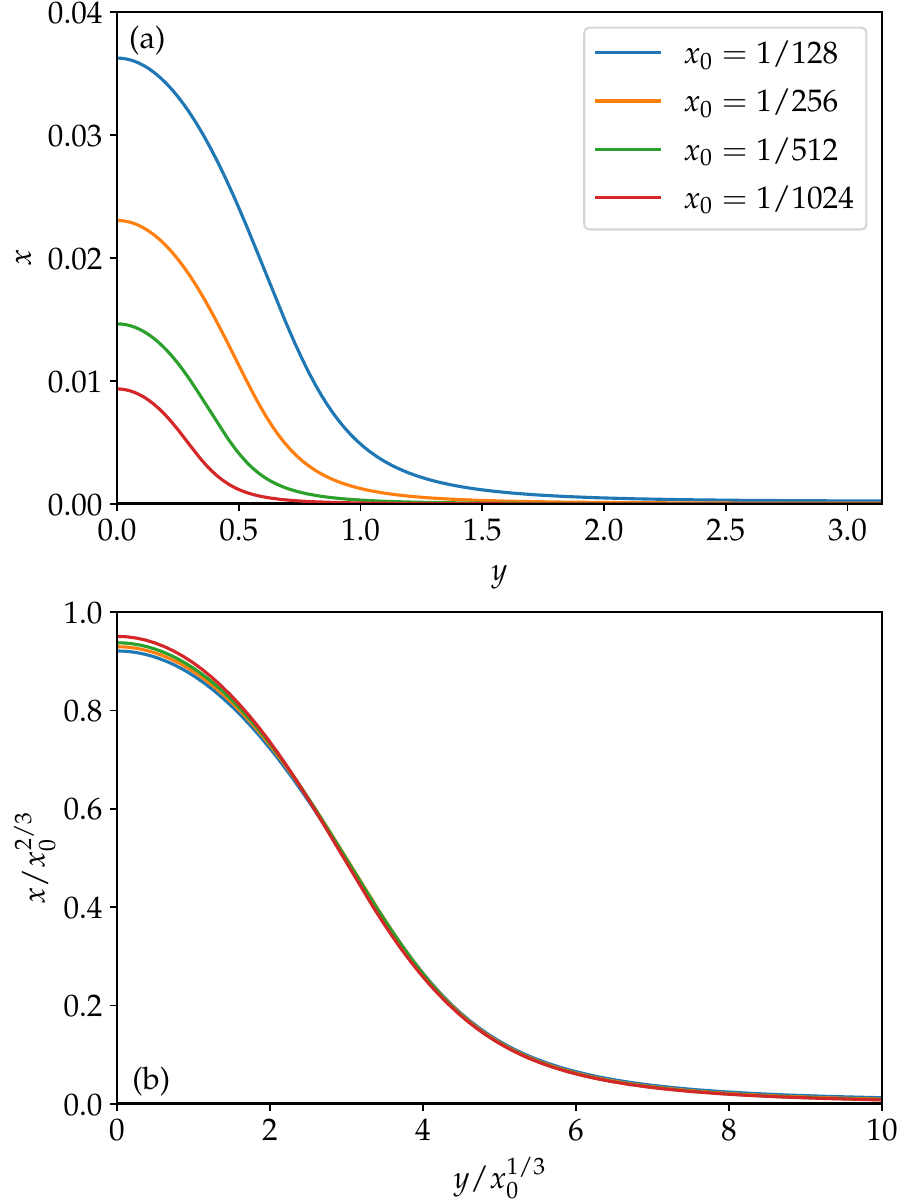}
\par\end{centering}
\caption{The nested flux surfaces near the resonant surface exhibit a similarity
relation. Panel (a) shows a selection of flux surfaces near the lower-left corner of Figure \ref{fig:flux-surfaces}. After rescaling by
$x\to x/x_{0}^{2/3}$ and $y\to y/x_{0}^{1/3}$, the flux surfaces
become nearly identical, as shown in panel (b). This figure uses the reference solution obtained by the GS solver with RDR subtraction.  Using the SPEC solution with packed volumes or the GS solver without RDR subtraction yields nearly identical curves.} \label{fig:similarity_relation}
\end{figure}

Without loss of generality, here we consider $x_{0}\ge0$. Applying
the leading order approximations of $f\left(x_{0}\right)$ and $g\left(y\right)$ near
$x_{0}=0$ and $y=0$ to the RDR solution (\ref{eq:RDR}) yields 

\begin{equation}
x\left(x_{0},y\right)\simeq\int_{0}^{x_{0}}\frac{x'}{\sqrt{d_{1}x'^{8/3}+d_{2}y^{8}}}dx',\label{eq:RDR-2}
\end{equation}
where $d_{1}$ and $d_{2}$ are some constants. With a change of variables
$\zeta=x'/x_{0}$, equation (\ref{eq:RDR-2}) can be rewritten as
\begin{equation}
\frac{x\left(x_{0},y\right)}{x_{0}^{2/3}}\simeq\int_{0}^{1}\frac{\zeta}{\sqrt{d_{1}\zeta^{8/3}+d_{2}\left(y/x_{0}^{1/3}\right)^{8}}}d\zeta.\label{eq:RDR-3}
\end{equation}
Equation (\ref{eq:RDR-3}) suggests that if we rescale $x$ and $y$
to $x/x_{0}^{2/3}$ and $y/x_{0}^{1/3}$, the flux surfaces near $(x,y)=(0,0)$
will approximately coincide. This similarity relation is borne out
by our numerical solutions, shown in Figure \ref{fig:similarity_relation}
for a selection of flux surfaces before and after rescaling. 

The similarity relation implies that the heights of the bulged
flux surfaces in the downstream region scale as $x_{0}^{2/3}$ and
the widths scale as $x_{0}^{1/3}$; the enclosed volumes scale as $x_{0}$, to be consistent with the incompressible constraint. Therefore,
in the limit of $x_{0}\to0$, the width of the bulged region becomes
narrower and narrower. Because the enclosed volumes scale as $\sim x_0$, the geometry of flux surfaces $x(x_0,y)$ may be viewed as approaching a Dirac $\delta$-function
$\sim x_{0}\delta(y)$.

\begin{figure}
\begin{centering}
\includegraphics[width=1\columnwidth]{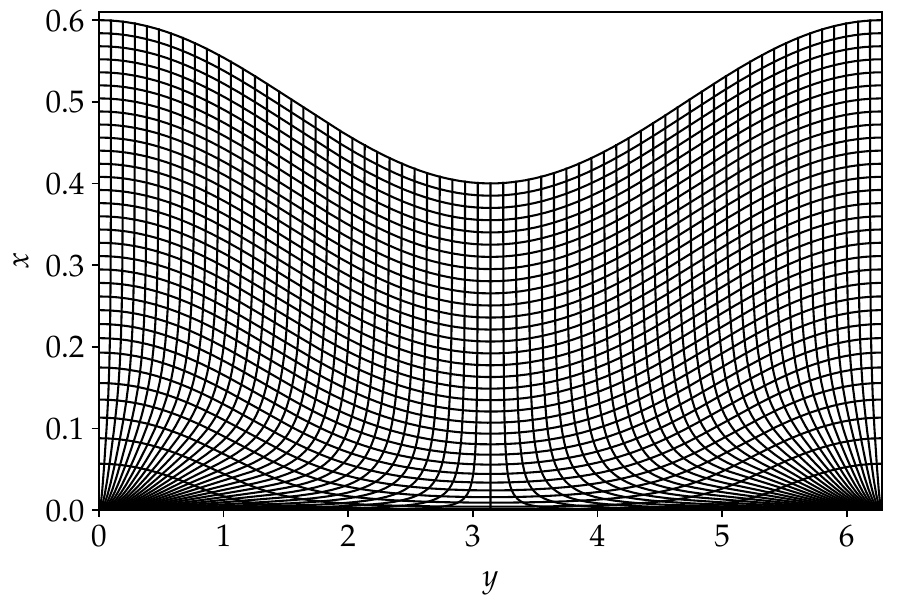}
\par\end{centering}
\caption{The distorted mesh shows the Lagrangian mapping to the final equilibrium
from a rectangular uniform mesh before imposing the boundary perturbation.\label{fig:Lagrangian_mapping}}
\end{figure}

Examining the solution from a Lagrangian perspective provides further
insight to its singular nature. The Lagrangian formulation of ideal
MHD describes a state in terms of the mapping from the initial positions
$\boldsymbol{x}_{0}$ of fluid elements to their final positions $\boldsymbol{x}$.
The magnetic field at $\boldsymbol{x}$ is determined by the initial
magnetic field $\boldsymbol{B}_{0}$ at $\boldsymbol{x}_{0}$ and
the mapping \textbf{$\boldsymbol{x}\left(\boldsymbol{x}_{0}\right)$
}via the relation \citep{Newcomb1962,ZhouQBB2014}
\begin{equation}
\boldsymbol{B}=\dfrac{\boldsymbol{B}_{0}\cdot\dfrac{\partial\boldsymbol{x}}{\partial\boldsymbol{x}_{0}}}{\mathcal{J}},\label{eq:B_lagrangian}
\end{equation}
where $\mathcal{J}=\det\left(\partial\boldsymbol{x}/\partial\boldsymbol{x}_{0}\right)$
is the Jacobian of the mapping. 

Although our GS solver is not fully Lagrangian because the mesh can
move along the $x$ direction but not along the $y$ direction, we
can reconstruct the full Lagrangian mapping of fluid elements from
the initial to the final state once the solution is obtained. For
each fluid element labeled by $\left(x_{0},y\right)$ in the final
solution, we need to find its initial position $\left(x_{0},y_{0}\right)$.
This ``inverse'' Lagrangian mapping can be expressed as a function
$y_{0}\left(x_{0},y\right)$. From the conservation of magnetic flux
through an infinitesimal fluid element 
\begin{equation}
B_{z0}\left(x_{0}\right)dx_{0}\left[\frac{\partial y_{0}}{\partial y}dy\right]=B_{z}\left(x_{0}\right)\left[\frac{\partial x}{\partial x_{0}}dx_{0}\right]dy\label{eq:conservation_flux}
\end{equation}
and using Eq.~(\ref{eq:Bz}) to relate $B_{z0}$ and $B_{z}$, we
can calculate 
\begin{equation}
\frac{\partial y_{0}}{\partial y}=\dfrac{\partial x/\partial x_{0}}{\left\langle \partial x/\partial x_{0}\right\rangle }\label{eq:dy0_dy}
\end{equation}
and integrate it along each constant-$x_{0}$ contour to obtain $y_{0}(x_{0},y)$.

Figure \ref{fig:Lagrangian_mapping} visualizes how a rectangular
uniform mesh in the initial state is deformed by the Lagrangian mapping
in the final state. We can see that the mapping is highly distorted
near the resonant surface. All the vertical mesh lines in the initial
state now converge towards the lower corners in the final state, and
the single point $\left(x_{0},y_{0}\right)=\left(0,L/2\right)$ is
stretched to an entire line of the lower boundary. This result strongly
suggests that the solution we find here for the HKT problem can only
be approached, but cannot be reached by ideal MHD evolution described
via smooth, diffeomorphic Lagrangian mapping.\citep{PfefferleNZ2020}

Now we discuss some limitations of our present methods in tackling the $\delta$-function singularities. At first sight, the pseudospectral method employed by the GS solver may seem ill-suited for problems with discontinuities. However, note that the primary variable to describe the solution is the geometry of flux surfaces represented by the mapping $x\left(x_0,y\right)$, which is not discontinuous. Although the magnetic field does become discontinuous, we only take derivatives on the total pressure $B^2/2+p$, which is continuous, when evaluating the residual force. For that reason, the pseudospectral method does not perform poorly because of the discontinuous magnetic field.  In this study, because we assume a mirror symmetry and solve for half of the domain, the magnetic field discontinuity is not present within the computational domain and therefore does not pose a problem. However, the GS solver works fine even when we do not assume the symmetry, provided that the collocation points do not fall on (or very close to) the resonant surface. 

Although the mapping $x\left(x_0,y\right)$ is continuous, it appears to become non-differentiable when the rotational transform is continuous. {Even though the exact form of $x\left(x_0,y\right)$ is not known, we may use the RDR solution, Eq.~(\ref{eq:RDR}), as a proxy. The function $x_{RDR}\left(x_0,y\right)$ is infinitely differentiable along the $y$ direction over the entire domain, and is infinitely differentiable along $x$ everywhere except at the point $(x_0, y)=(0,0)$ (and also $(0,L)$ because of the periodicity). Because $x \sim x_0^{2/3}$ when $y=0$, the partial derivative of $x_{RDR}$ along $x_0$ diverges at $(x_0, y)=(0,0)$ as $\partial x_{RDR}/\partial x_0|_{y=0} \sim x_0^{-1/3}$. This singular behavior leads to the non-smoothness of the flux surfaces near the resonant surface.} Consequently, the convergence rate of the GS solver is algebraic with respect to the number of collocation points (see Figures \ref{fig:Convergence-of-By-err} and \ref{fig:Convergence-of-flux-surface}), as opposed to an exponential convergence we usually expect from a pseudospectral method. In contrast, when applying to a problem that satisfies the \emph{sine qua non} condition in Sec.~\ref{sec:sine_qua_non}, the GS solver can achieve much more rapid convergence (see Figures \ref{fig:By_err_discontinuous_iota} and \ref{fig:surface_err_discontinuous_iota}).

We can appreciate the non-smoothness of HKT flux surfaces near the resonant surface through the similarity relation we discussed earlier. Because the width of the bulged region scales as $x_0^{1/3}$, when we increase the resolution along the $x$ direction,  we need to increase the resolution along $y$ direction as well to resolve the localized structure. For the GS solver, since the closest Chebyshev collocation point to the resonant surface has $x_0 \propto 1/N_x^2$, roughly speaking, the resolution in $y$ needs to scale as $N_y \sim N_x^{2/3}$ to resolve the localized structures. A similar requirement also applies to SPEC when the number of volumes increases.
Therefore, the Fourier representation employed by both solvers is inefficient for ideal flux surfaces near the resonant surface. A possible remedy is to employ alternative basis functions for the flux surfaces. The version of GS solver with RDR subtraction effectively uses the RDR solution as one of the basis functions. However, although subtracting the RDR solution significantly improves the accuracy of the GS solver, it does not completely remove the effect of the singularity and the convergence rate remains similarly algebraic. The convergence rate could potentially be further improved by adopting a more accurate $g\left(y\right)$ in the RDR solution (\ref{eq:RDR}), either by numerically solving the RDR integral equation \citep{LoizuH2017} or by dynamically solving $g\left(y\right)$ as a part of the solver.

Note that this singular behavior of flux surfaces near the resonant surface only arises when we try to obtain the ideal MHD solution. SPEC, which implements MRxMHD, is not an ideal MHD equilibrium solver with nested flux surfaces by design. By changing the number of volumes, SPEC allows a transition from Taylor relaxation to ideal MHD.  {When modeling non-ideal plasmas that allow magnetic islands and regions of stochastic field lines with MRxMHD}, an active area of research is to understand where the ideal interfaces should be placed and when an ideal interface should be removed.\citep{QuHDLH2021} The presence of a strong current sheet on an ideal interface is an indication that the interface should be removed. If we remove the ideal interface at $x_0=0$ and allow reconnection, the singular behavior of flux surfaces may no longer be a problem.

\subsection{Reinterpreting the \emph{sine qua non} condition in Loizu \emph{et al.} (2015)\citep{LoizuHBLH2015} \label{sec:sine_qua_non}}

We mention in the Introduction that Loizu \emph{et al.}\citep{LoizuHBLH2015} previously studied an $m=2$, $n=1$
perturbation on a cylindrical screw pinch with SPEC and concluded that a minimal finite jump in the rotational transform is necessary for the existence of a solution. This finding motivated Loizu \emph{et al.} to call the minimal finite jump a \emph{sine qua non} condition. However, in the present study, we show that SPEC actually can find solutions for the HKT problem, which has a continuous rotational transform, provided that Newton's method is initialized with care. Therefore, the previous interpretation of the \emph{sine qua non} condition by Loizu \emph{et al.} is incorrect. To further clarify the issue, it is instructive to discuss the \emph{sine qua non} condition in the context of the HKT problem. A similar discussion can also be found in Sec.~3.3 of Ref.~[\onlinecite{Zhou2017}].

Instead of a continuous initial magnetic field, let us now suppose that the initial field has a finite discontinuity at $x_0=0$:
\begin{equation}
    B_{y0}=x_{0}\pm b.
    \label{eq: discontinous_B_init}
\end{equation}
Here, we take the plus sign for $x_0>0$ and the minus sign for $x_0<0$. The discontinuity parameter $b$ provides a finite jump in the rotational transform. In the limit $b\to 0$, the original HKT problem is recovered.

For this modified HKT problem, the linearized GS equation (\ref{eq:GS-linear}) becomes 
\begin{equation}
\frac{d^{2}}{dx_{0}^{2}}\left(\left(x_{0}\pm b\right)\bar{\xi}\right)-k^{2}\left(x_{0}\pm b\right)\bar{\xi}=0.
\label{eq:GS-linear_HKT_b}
\end{equation}
With boundary conditions $\bar{\xi}(0)=0$ and $\bar{\xi}(\pm a)=\pm\delta$, the solution is \begin{equation}
\bar{\xi}=\frac{(a+b)\delta}{\sinh(ka)}\frac{\sinh\left(kx_{0}\right)}{x_{0}\pm b}.
\label{eq:linear_sol_b}
\end{equation}

The geometry of perturbed flux surfaces up to the linear order is given by $x=x_0+\bar\xi \cos(ky)$. To prevent overlapping of flux surfaces requires $\partial x/\partial x_0>0$, which amounts to
\begin{equation}
    \left|\frac{d\bar{\xi}}{dx_{0}}\right|=  \frac{(a+b)\delta}{\sinh(ka)}\left|k\frac{\cosh\left(kx_{0}\right)}{x_{0}\pm b}-\frac{\sinh\left(kx_{0}\right)}{(x_{0} \pm b)^{2}}\right|<1
\end{equation}
for the linear perturbation. Since  
\begin{equation}
    \left|\frac{d\bar{\xi}}{dx_{0}}\right|=  \frac{(a+b)k\delta}{b\sinh(ka)}\left|1\mp\frac {2x_0}{b}+O\left(x_0^2\right)\right|
\end{equation}
in the vicinity of $x_0=0$, it is sufficient to ensure that $\left|d\bar{\xi}/dx_0\right|<1$ at $x_0=0$. That leads to the \emph{sine qua non} condition for the HKT problem:
\begin{equation}
    b>b_{\min}=\frac{ka\delta}{\sinh\left(ka\right)-k\delta}.\label{eq:bmin}
\end{equation}

The \emph{sine qua non} condition ensures that the flux surfaces of the linear solution do not overlap.  However, not satisfying the \emph{sine qua non} condition does not imply the nonexistence of a solution; it simply means that a nonlinear solution must be sought non-perturbatively.  The RDR solution demonstrates how an approximate nonlinear solution can be obtained through a boundary layer analysis and asymptotic matching.  The previous misinterpretation of the \emph{sine qua non} condition  as the necessary condition for the existence of a solution  further led to an erroneous claim that the RDR solution has a discontinuous rotational transform.\citep{LoizuH2017} This latter mistake has been corrected by Zhou \emph{et al.}\citep{ZhouHRQB2019}

Let us now examine how the \emph{sine qua non} condition affects the solution. For the same boundary perturbation with $k=1$, $a=0.5$, and $\delta=0.1$ as before, the \emph{sine qua non} condition (\ref{eq:bmin}) gives $b_{\min}\simeq 0.119$. In what follows, we consider the case $b=0.12$ such that the \emph{sine qua non} condition is marginally satisfied. We numerically calculate the solution with the GS solver and perform exactly the same convergence tests as before.

\begin{figure}
\begin{centering}
\includegraphics[width=1\columnwidth]{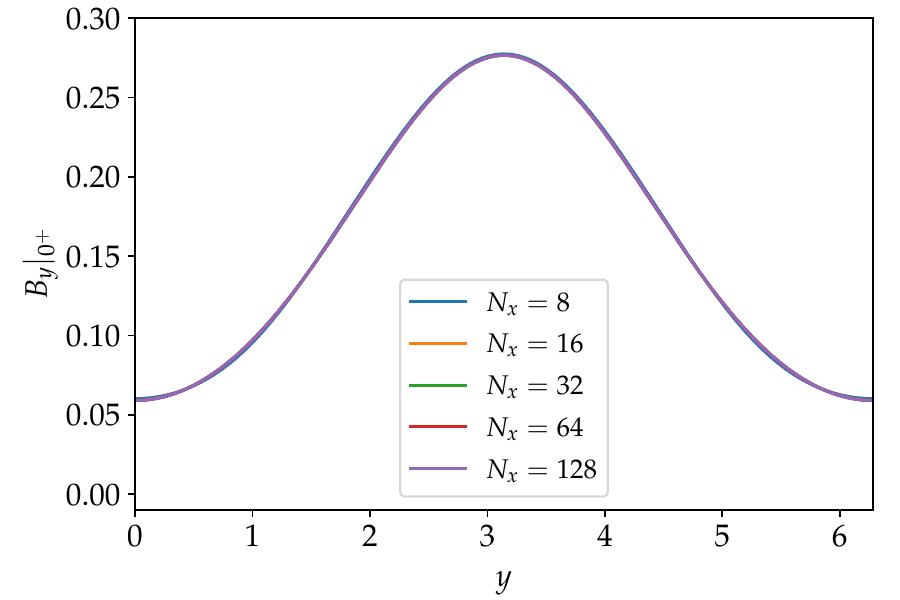}
\par\end{centering}
\caption{The magnetic field discontinuity $B_y|_{0^+}$ obtained by the GS solver with different number of collocation points $N_x$, for the case $b=0.12$. } \label{fig:By_discontinuous_iota}
\end{figure}

\begin{figure}
\begin{centering}
\includegraphics[width=1\columnwidth]{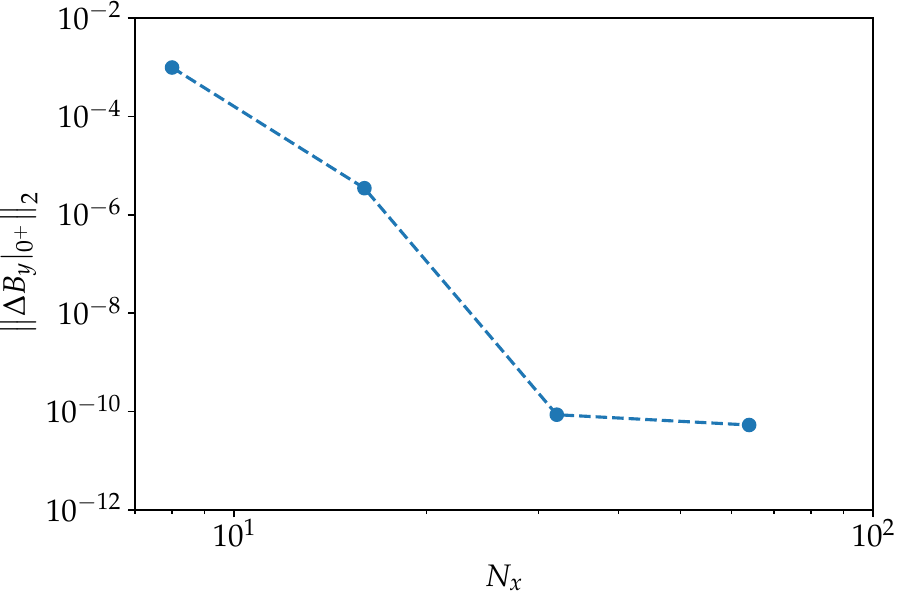}
\par\end{centering}
\caption{Self convergence test of the $B_y|_{0^+}$ error for the case $b=0.12$. The $N_x=128$ solution serves as the reference. The error is dominated by round-off errors when $N_x\ge32$.} \label{fig:By_err_discontinuous_iota}
\end{figure}

\begin{figure}
\begin{centering}
\includegraphics[width=1\columnwidth]{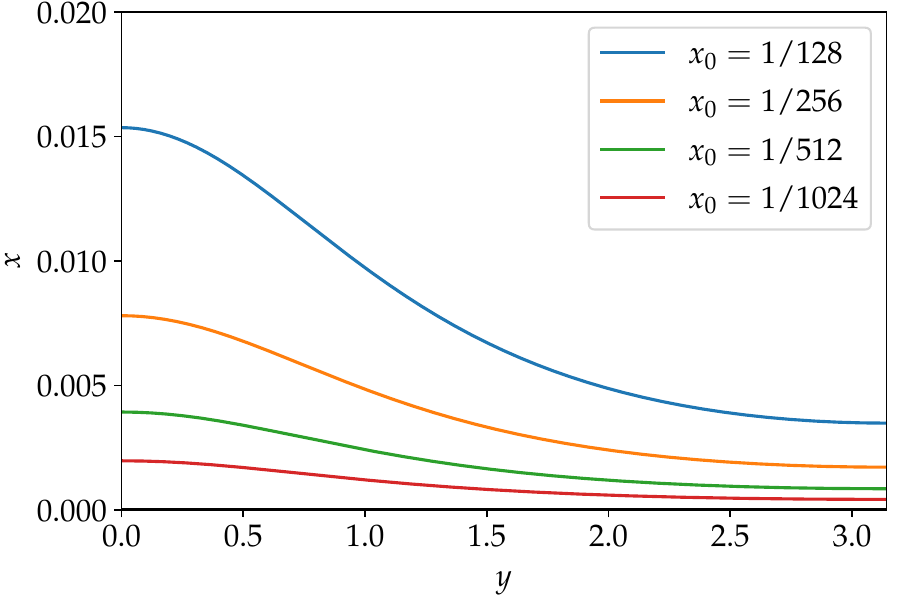}
\par\end{centering}
\caption{A zoom-in view of a selection of flux surfaces for the case $b=0.12$. This figure can be compared with Fig.~\ref{fig:similarity_relation}(a) for the case $b=0$.} \label{fig:flux_surfaces_discontinuous_iota_zoom}
\end{figure}

Figure \ref{fig:By_discontinuous_iota} shows the magnetic field discontinuity $B_y|_{0^+}$ for different number of collocation points $N_x$. We can see that all the curves are visually indistinguishable, even with a resolution as low as $N_x=8$. Note that $B_y|_{0^+}>0$ everywhere because of the discontinuous rotational transform.

Figure \ref{fig:By_err_discontinuous_iota} shows a self convergence test for the $B_y|_{0^+}$ error. Here, we use the $N_x=128$ solution as the reference. We can see that the error reaches a level below $10^{-10}$ at $N_x=32$. Further increasing the resolution does not lower the error, suggesting that the error is dominated by round-off errors when $N_x\ge32$.

\begin{figure}
\begin{centering}
\includegraphics[width=1\columnwidth]{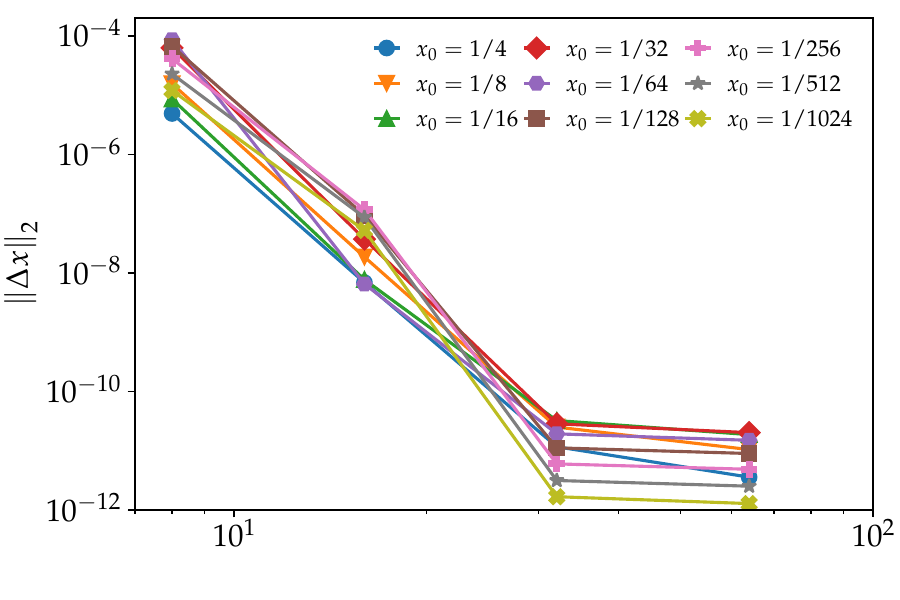}
\par\end{centering}
\caption{Self convergence test of the flux surface errors for the case $b=0.12$. The $N_x=128$ solution serves as the reference. The errors are dominated by round-off errors when $N_x\ge32$.}  \label{fig:surface_err_discontinuous_iota}
\end{figure}

Figure \ref{fig:flux_surfaces_discontinuous_iota_zoom} shows a zoom-in view of a selection of flux surfaces. These flux surfaces correspond to the same flux surfaces shown in Fig.~\ref{fig:similarity_relation}(a) for the case with a continuous rotational transform. Comparing these two figures, we can see that with the discontinuous rotational transform, the space between flux surfaces is no longer strongly squeezed; consequently, the flux surfaces also do not  strongly bulge out in the ``downstream'' region near the lower-left corner. Figure \ref{fig:surface_err_discontinuous_iota} shows a self convergence test of flux surface errors, using the $N_x=128$ case as the reference. Again, the errors appear to be limited by round-off errors when $N_x\ge32$.

The numerical calculations shown here all use $N_y=512$. However, because the flux surfaces no longer have strongly localized geometric structures, the same accuracy can be achieved with a lot fewer grid points along $y$. We find that a similar accuracy can be achieved with $N_x=32$ and $N_y=64$. 

This example demonstrates that the \emph{sine qua non} condition significantly alters the nature of the solution, which may have contributed to why SPEC had no problem finding solutions when the condition was satisfied in Ref.~[\onlinecite{LoizuHBLH2015}]. Because the space between flux surfaces is no longer strongly squeezed, the Newton's solver of SPEC is less likely to have overlapped ideal interfaces. Moreover, because the SPEC calculations in Ref.~[\onlinecite{LoizuHBLH2015}] only use a small number of Fourier harmonics (the toroidal mode number $n\le6$ and the poloidal mode number $m\le3$), the fact that the flux surfaces do not develop localized structures also helps.



\section{Conclusions and Future Perspectives\label{sec:Conclusions}}

In conclusion, we have demonstrated that with the increase of resolution
or the number of volumes, the GS solver and SPEC both
appear to approach  the solution of the ideal HKT problem with a $\delta$-function singularity.
Our result is also the first to show that SPEC can obtain approximate
solutions of the ideal HKT problem without requiring a discontinuous rotational
transform across the resonant surface in the initial condition. 

In the previous calculation by Loizu {\it et al.},\citep{LoizuHBLH2015} the {\it sine qua non} condition originated from a breakdown of the linear solution near the resonant surface, which was misinterpreted as a lack of a solution. This misunderstanding, compounded with the fact that SPEC uses a Newton method that may fail to find the solution without a carefully chosen initial guess, led to the erroneous conclusion that a finite threshold of discontinuous rotational transform is necessary for the existence of a solution. However, as we have discussed in Sec.~\ref{sec:Grad-Shafranov-solutions}, the breakdown of the linear solution does not imply a lack of a solution, but rather that a nonlinear solution must be sought. Furthermore, by carefully initiating Newton's method, we have demonstrated that SPEC can obtain the solution. The present study also calls for reconsideration of  the previous study by Loizu {\it et al.},\citep{LoizuHBLH2015} as well as the benchmark study between SPEC and VMEC on the same problem. \citep{LazersonLHH2016} 

In future work, it would be prudent to implement in SPEC a steepest descent algorithm for the energy functional, which should be beneficial for tackling this and similar problems. For instance, one could first use the more robust steepest descent algorithm to obtain an approximate solution, then switch to Newton's method for more rapid convergence to the final solution.

Following up this work, several further investigations will be pursued in the future. Some of the present approaches could be adapted to singular current sheets arising from the ideal internal kink instability \citep{RosenbluthDR1973,ParkMWJ1980} and more general 3D magnetic resonant perturbations. In addition to the $\delta$-function singularities, the algebraically divergent Pfirsch--Schl\"uter current in the presence of a pressure gradient should also be investigated. Recent studies have shown that ideal current singularities on resonant surfaces may be eliminated by  modifying the plasma boundary.\citep{MikhailovNZ2019, KimMC2020} This new approach could also be investigated with SPEC. Finally, the tendency to form current sheets is thwarted in real plasmas by non-ideal effects, which will tend to drive magnetic reconnection, forming magnetic islands or regions of stochastic field
lines when island overlap occurs. An important question of practical significance is whether
the sizes of saturated islands or regions of stochastic field lines
can be predicted from the intensity of current singularities. \citep{CaryH1991, BhattacharjeeHHNS1995,LoizuHHBKQ2020,GeraldiniLP2021,RodriguezB2021} If such a relationship can be established, we may use singularity intensities as a proxy for the sizes of magnetic islands (or regions of stochastic field lines) in stellarator optimization.

\begin{acknowledgments}
This research was supported by the U.S. Department of Energy under Contract No. DE-AC02-09CH11466 and by a grant from the Simons Foundation/SFARI (560651, AB). Part of this work has been carried out within the framework of the EUROfusion Consortium and has received funding from the Euratom research and training programme 2014–2018 and 2019–2020 under Grant Agreement No. 633053. The views and opinions expressed herein do not necessarily reflect those of the European Commission. YZ was partly sponsored by Shanghai Pujiang Program under Grant No. 21PJ1408600. Part of the numerical calculations were performed with computers at the National Energy Research Scientific Computing Center.
\end{acknowledgments}

\section*{Data Availability}
The data that support the findings of this study are available
from the corresponding author upon reasonable request.

\bibliographystyle{apsrev4-1}
\bibliography{reference}

\begin{thebibliography}{39}%
\makeatletter
\providecommand \@ifxundefined [1]{%
 \@ifx{#1\undefined}
}%
\providecommand \@ifnum [1]{%
 \ifnum #1\expandafter \@firstoftwo
 \else \expandafter \@secondoftwo
 \fi
}%
\providecommand \@ifx [1]{%
 \ifx #1\expandafter \@firstoftwo
 \else \expandafter \@secondoftwo
 \fi
}%
\providecommand \natexlab [1]{#1}%
\providecommand \enquote  [1]{``#1''}%
\providecommand \bibnamefont  [1]{#1}%
\providecommand \bibfnamefont [1]{#1}%
\providecommand \citenamefont [1]{#1}%
\providecommand \href@noop [0]{\@secondoftwo}%
\providecommand \href [0]{\begingroup \@sanitize@url \@href}%
\providecommand \@href[1]{\@@startlink{#1}\@@href}%
\providecommand \@@href[1]{\endgroup#1\@@endlink}%
\providecommand \@sanitize@url [0]{\catcode `\\12\catcode `\$12\catcode
  `\&12\catcode `\#12\catcode `\^12\catcode `\_12\catcode `\%12\relax}%
\providecommand \@@startlink[1]{}%
\providecommand \@@endlink[0]{}%
\providecommand \url  [0]{\begingroup\@sanitize@url \@url }%
\providecommand \@url [1]{\endgroup\@href {#1}{\urlprefix }}%
\providecommand \urlprefix  [0]{URL }%
\providecommand \Eprint [0]{\href }%
\providecommand \doibase [0]{http://dx.doi.org/}%
\providecommand \selectlanguage [0]{\@gobble}%
\providecommand \bibinfo  [0]{\@secondoftwo}%
\providecommand \bibfield  [0]{\@secondoftwo}%
\providecommand \translation [1]{[#1]}%
\providecommand \BibitemOpen [0]{}%
\providecommand \bibitemStop [0]{}%
\providecommand \bibitemNoStop [0]{.\EOS\space}%
\providecommand \EOS [0]{\spacefactor3000\relax}%
\providecommand \BibitemShut  [1]{\csname bibitem#1\endcsname}%
\let\auto@bib@innerbib\@empty
\bibitem [{\citenamefont {Parker}(1994)}]{Parker1994}%
  \BibitemOpen
  \bibfield  {author} {\bibinfo {author} {\bibfnamefont {E.~N.}\ \bibnamefont
  {Parker}},\ }\href@noop {} {\emph {\bibinfo {title} {Spontaneous Current
  Sheets in Magnetic Fields}}}\ (\bibinfo  {publisher} {Oxford University
  Press, Inc.},\ \bibinfo {year} {1994})\BibitemShut {NoStop}%
\bibitem [{\citenamefont {Hirshman}\ and\ \citenamefont
  {Whitson}(1983)}]{HirshmanW1983}%
  \BibitemOpen
  \bibfield  {author} {\bibinfo {author} {\bibfnamefont {S.~P.}\ \bibnamefont
  {Hirshman}}\ and\ \bibinfo {author} {\bibfnamefont {J.~C.}\ \bibnamefont
  {Whitson}},\ }\href {\doibase 10.1063/1.864116} {\bibfield  {journal}
  {\bibinfo  {journal} {Physics of Fluids}\ }\textbf {\bibinfo {volume} {26}},\
  \bibinfo {pages} {3553} (\bibinfo {year} {1983})}\BibitemShut {NoStop}%
\bibitem [{\citenamefont {Garabedian}(2002)}]{Garabedian2002}%
  \BibitemOpen
  \bibfield  {author} {\bibinfo {author} {\bibfnamefont {P.~R.}\ \bibnamefont
  {Garabedian}},\ }\href {\doibase 10.1073/pnas.162330399} {\bibfield
  {journal} {\bibinfo  {journal} {Proceedings of the National Academy of
  Sciences}\ }\textbf {\bibinfo {volume} {99}},\ \bibinfo {pages} {10257}
  (\bibinfo {year} {2002})}\BibitemShut {NoStop}%
\bibitem [{\citenamefont {Grad}(1967)}]{Grad1967}%
  \BibitemOpen
  \bibfield  {author} {\bibinfo {author} {\bibfnamefont {H.}~\bibnamefont
  {Grad}},\ }\href@noop {} {\bibfield  {journal} {\bibinfo  {journal} {Phys.
  Fluids}\ }\textbf {\bibinfo {volume} {10}},\ \bibinfo {pages} {137} (\bibinfo
  {year} {1967})}\BibitemShut {NoStop}%
\bibitem [{\citenamefont {Bhattacharjee}\ \emph {et~al.}(1995)\citenamefont
  {Bhattacharjee}, \citenamefont {Hayashi}, \citenamefont {Hegna},
  \citenamefont {Nakajima},\ and\ \citenamefont
  {Sato}}]{BhattacharjeeHHNS1995}%
  \BibitemOpen
  \bibfield  {author} {\bibinfo {author} {\bibfnamefont {A.}~\bibnamefont
  {Bhattacharjee}}, \bibinfo {author} {\bibfnamefont {T.}~\bibnamefont
  {Hayashi}}, \bibinfo {author} {\bibfnamefont {C.~C.}\ \bibnamefont {Hegna}},
  \bibinfo {author} {\bibfnamefont {N.}~\bibnamefont {Nakajima}}, \ and\
  \bibinfo {author} {\bibfnamefont {T.}~\bibnamefont {Sato}},\ }\href@noop {}
  {\bibfield  {journal} {\bibinfo  {journal} {Phys. Plasmas}\ }\textbf
  {\bibinfo {volume} {2}},\ \bibinfo {pages} {883} (\bibinfo {year}
  {1995})}\BibitemShut {NoStop}%
\bibitem [{\citenamefont {Helander}(2014)}]{Helander2014}%
  \BibitemOpen
  \bibfield  {author} {\bibinfo {author} {\bibfnamefont {P.}~\bibnamefont
  {Helander}},\ }\href {\doibase 10.1088/0034-4885/77/8/087001} {\bibfield
  {journal} {\bibinfo  {journal} {Reports on Progress in Physics}\ }\textbf
  {\bibinfo {volume} {77}},\ \bibinfo {pages} {087001} (\bibinfo {year}
  {2014})}\BibitemShut {NoStop}%
\bibitem [{\citenamefont {Biskamp}(1993)}]{Biskamp1993}%
  \BibitemOpen
  \bibfield  {author} {\bibinfo {author} {\bibfnamefont {D.}~\bibnamefont
  {Biskamp}},\ }\href@noop {} {\emph {\bibinfo {title} {Nonlinear
  Magnetohydrodynamics}}}\ (\bibinfo  {publisher} {Cambridge University
  Press},\ \bibinfo {year} {1993})\BibitemShut {NoStop}%
\bibitem [{\citenamefont {Biskamp}(2000)}]{Biskamp2000}%
  \BibitemOpen
  \bibfield  {author} {\bibinfo {author} {\bibfnamefont {D.}~\bibnamefont
  {Biskamp}},\ }\href@noop {} {\emph {\bibinfo {title} {Magnetic Reconnection
  in Plasmas}}}\ (\bibinfo  {publisher} {Cambridge University Press},\ \bibinfo
  {year} {2000})\BibitemShut {NoStop}%
\bibitem [{\citenamefont {Hahm}\ and\ \citenamefont
  {Kulsrud}(1985)}]{HahmK1985}%
  \BibitemOpen
  \bibfield  {author} {\bibinfo {author} {\bibfnamefont {T.~S.}\ \bibnamefont
  {Hahm}}\ and\ \bibinfo {author} {\bibfnamefont {R.~M.}\ \bibnamefont
  {Kulsrud}},\ }\href@noop {} {\bibfield  {journal} {\bibinfo  {journal} {Phys.
  Fluids}\ }\textbf {\bibinfo {volume} {28}},\ \bibinfo {pages} {2412}
  (\bibinfo {year} {1985})}\BibitemShut {NoStop}%
\bibitem [{\citenamefont {Wang}\ and\ \citenamefont
  {Bhattacharjee}(1992)}]{WangB1992a}%
  \BibitemOpen
  \bibfield  {author} {\bibinfo {author} {\bibfnamefont {X.}~\bibnamefont
  {Wang}}\ and\ \bibinfo {author} {\bibfnamefont {A.}~\bibnamefont
  {Bhattacharjee}},\ }\href@noop {} {\bibfield  {journal} {\bibinfo  {journal}
  {Phys. Fluids B}\ }\textbf {\bibinfo {volume} {4}},\ \bibinfo {pages} {1795}
  (\bibinfo {year} {1992})}\BibitemShut {NoStop}%
\bibitem [{\citenamefont {Dewar}\ \emph {et~al.}(2017)\citenamefont {Dewar},
  \citenamefont {Hudson}, \citenamefont {Bhattacharjee},\ and\ \citenamefont
  {Yoshida}}]{DewarHBY2017}%
  \BibitemOpen
  \bibfield  {author} {\bibinfo {author} {\bibfnamefont {R.~L.}\ \bibnamefont
  {Dewar}}, \bibinfo {author} {\bibfnamefont {S.~R.}\ \bibnamefont {Hudson}},
  \bibinfo {author} {\bibfnamefont {A.}~\bibnamefont {Bhattacharjee}}, \ and\
  \bibinfo {author} {\bibfnamefont {Z.}~\bibnamefont {Yoshida}},\ }\href
  {\doibase 10.1063/1.4979350} {\bibfield  {journal} {\bibinfo  {journal}
  {Phys. Plasmas}\ }\textbf {\bibinfo {volume} {24}},\ \bibinfo {pages}
  {042507} (\bibinfo {year} {2017})}\BibitemShut {NoStop}%
\bibitem [{\citenamefont {Huang}\ \emph {et~al.}(2009)\citenamefont {Huang},
  \citenamefont {Bhattacharjee},\ and\ \citenamefont {Zweibel}}]{HuangBZ2009}%
  \BibitemOpen
  \bibfield  {author} {\bibinfo {author} {\bibfnamefont {Y.-M.}\ \bibnamefont
  {Huang}}, \bibinfo {author} {\bibfnamefont {A.}~\bibnamefont
  {Bhattacharjee}}, \ and\ \bibinfo {author} {\bibfnamefont {E.~G.}\
  \bibnamefont {Zweibel}},\ }\href {\doibase 10.1088/0004-637X/699/2/L144}
  {\bibfield  {journal} {\bibinfo  {journal} {Astrophys. J. Lett.}\ }\textbf
  {\bibinfo {volume} {699}},\ \bibinfo {pages} {L144} (\bibinfo {year}
  {2009})}\BibitemShut {NoStop}%
\bibitem [{\citenamefont {Hudson}\ \emph {et~al.}(2012)\citenamefont {Hudson},
  \citenamefont {Dewar}, \citenamefont {Dennis}, \citenamefont {Hole},
  \citenamefont {McGann}, \citenamefont {von Nessi},\ and\ \citenamefont
  {Lazerson}}]{HudsonDDHMNL2012}%
  \BibitemOpen
  \bibfield  {author} {\bibinfo {author} {\bibfnamefont {S.~R.}\ \bibnamefont
  {Hudson}}, \bibinfo {author} {\bibfnamefont {R.~L.}\ \bibnamefont {Dewar}},
  \bibinfo {author} {\bibfnamefont {G.}~\bibnamefont {Dennis}}, \bibinfo
  {author} {\bibfnamefont {M.~J.}\ \bibnamefont {Hole}}, \bibinfo {author}
  {\bibfnamefont {M.}~\bibnamefont {McGann}}, \bibinfo {author} {\bibfnamefont
  {G.}~\bibnamefont {von Nessi}}, \ and\ \bibinfo {author} {\bibfnamefont
  {S.}~\bibnamefont {Lazerson}},\ }\href {\doibase 10.1063/1.4765691}
  {\bibfield  {journal} {\bibinfo  {journal} {Physics of Plasmas}\ }\textbf
  {\bibinfo {volume} {19}},\ \bibinfo {pages} {112502} (\bibinfo {year}
  {2012})}\BibitemShut {NoStop}%
\bibitem [{\citenamefont {Fornberg}(1995)}]{Fornberg1995a}%
  \BibitemOpen
  \bibfield  {author} {\bibinfo {author} {\bibfnamefont {B.}~\bibnamefont
  {Fornberg}},\ }\href@noop {} {\emph {\bibinfo {title} {A Practical Guide to
  Pseudospectral Methods}}}\ (\bibinfo  {publisher} {Cambridge University
  Press},\ \bibinfo {year} {1995})\BibitemShut {NoStop}%
\bibitem [{\citenamefont {Trefethen}(2000)}]{Trefethen2000}%
  \BibitemOpen
  \bibfield  {author} {\bibinfo {author} {\bibfnamefont {L.~N.}\ \bibnamefont
  {Trefethen}},\ }\href@noop {} {\emph {\bibinfo {title} {Spectral Methods in
  Matlab}}}\ (\bibinfo  {publisher} {SIAM Philadelphia},\ \bibinfo {year}
  {2000})\BibitemShut {NoStop}%
\bibitem [{\citenamefont {Zhou}\ \emph {et~al.}(2016)\citenamefont {Zhou},
  \citenamefont {Huang}, \citenamefont {Qin},\ and\ \citenamefont
  {Bhattacharjee}}]{ZhouHQB2016}%
  \BibitemOpen
  \bibfield  {author} {\bibinfo {author} {\bibfnamefont {Y.}~\bibnamefont
  {Zhou}}, \bibinfo {author} {\bibfnamefont {Y.-M.}\ \bibnamefont {Huang}},
  \bibinfo {author} {\bibfnamefont {H.}~\bibnamefont {Qin}}, \ and\ \bibinfo
  {author} {\bibfnamefont {A.}~\bibnamefont {Bhattacharjee}},\ }\href {\doibase
  10.1103/PhysRevE.93.023205} {\bibfield  {journal} {\bibinfo  {journal} {Phys.
  Rev. E}\ }\textbf {\bibinfo {volume} {93}},\ \bibinfo {pages} {023205}
  (\bibinfo {year} {2016})}\BibitemShut {NoStop}%
\bibitem [{\citenamefont {Zhou}\ \emph {et~al.}(2019)\citenamefont {Zhou},
  \citenamefont {Huang}, \citenamefont {Reiman}, \citenamefont {Qin},\ and\
  \citenamefont {Bhattacharjee}}]{ZhouHRQB2019}%
  \BibitemOpen
  \bibfield  {author} {\bibinfo {author} {\bibfnamefont {Y.}~\bibnamefont
  {Zhou}}, \bibinfo {author} {\bibfnamefont {Y.-M.}\ \bibnamefont {Huang}},
  \bibinfo {author} {\bibfnamefont {A.~H.}\ \bibnamefont {Reiman}}, \bibinfo
  {author} {\bibfnamefont {H.}~\bibnamefont {Qin}}, \ and\ \bibinfo {author}
  {\bibfnamefont {A.}~\bibnamefont {Bhattacharjee}},\ }\href {\doibase
  10.1063/1.5068778} {\bibfield  {journal} {\bibinfo  {journal} {Physics of
  Plasmas}\ }\textbf {\bibinfo {volume} {26}},\ \bibinfo {pages} {022103}
  (\bibinfo {year} {2019})}\BibitemShut {NoStop}%
\bibitem [{\citenamefont {Taylor}(1974)}]{Taylor1974}%
  \BibitemOpen
  \bibfield  {author} {\bibinfo {author} {\bibfnamefont {J.~B.}\ \bibnamefont
  {Taylor}},\ }\href@noop {} {\bibfield  {journal} {\bibinfo  {journal} {Phys.
  Rev. Lett.}\ }\textbf {\bibinfo {volume} {33}},\ \bibinfo {pages} {1139}
  (\bibinfo {year} {1974})}\BibitemShut {NoStop}%
\bibitem [{Note1()}]{Note1}%
  \BibitemOpen
  \bibinfo {note} {However, note that stochastic field line regions are not
  possible for the 2D HKT problem even when magnetic reconnection is allowed;
  only magnetic islands are possible.}\BibitemShut {Stop}%
\bibitem [{\citenamefont {Dennis}\ \emph {et~al.}(2013)\citenamefont {Dennis},
  \citenamefont {Hudson}, \citenamefont {Dewar},\ and\ \citenamefont
  {Hole}}]{DennisHDH2013}%
  \BibitemOpen
  \bibfield  {author} {\bibinfo {author} {\bibfnamefont {G.~R.}\ \bibnamefont
  {Dennis}}, \bibinfo {author} {\bibfnamefont {S.~R.}\ \bibnamefont {Hudson}},
  \bibinfo {author} {\bibfnamefont {R.~L.}\ \bibnamefont {Dewar}}, \ and\
  \bibinfo {author} {\bibfnamefont {M.~J.}\ \bibnamefont {Hole}},\ }\href
  {\doibase 10.1063/1.4795739} {\bibfield  {journal} {\bibinfo  {journal}
  {Physics of Plasmas}\ }\textbf {\bibinfo {volume} {20}},\ \bibinfo {pages}
  {032509} (\bibinfo {year} {2013})}\BibitemShut {NoStop}%
\bibitem [{\citenamefont {Qu}\ \emph {et~al.}(2021)\citenamefont {Qu},
  \citenamefont {Hudson}, \citenamefont {Dewar}, \citenamefont {Loizu},\ and\
  \citenamefont {Hole}}]{QuHDLH2021}%
  \BibitemOpen
  \bibfield  {author} {\bibinfo {author} {\bibfnamefont {Z.~S.}\ \bibnamefont
  {Qu}}, \bibinfo {author} {\bibfnamefont {S.~R.}\ \bibnamefont {Hudson}},
  \bibinfo {author} {\bibfnamefont {R.~L.}\ \bibnamefont {Dewar}}, \bibinfo
  {author} {\bibfnamefont {J.}~\bibnamefont {Loizu}}, \ and\ \bibinfo {author}
  {\bibfnamefont {M.~J.}\ \bibnamefont {Hole}},\ }\href {\doibase
  10.1088/1361-6587/ac2afc} {\bibfield  {journal} {\bibinfo  {journal} {Plasma
  Physics and Controlled Fusion}\ }\textbf {\bibinfo {volume} {63}},\ \bibinfo
  {pages} {125007} (\bibinfo {year} {2021})}\BibitemShut {NoStop}%
\bibitem [{\citenamefont {Loizu}\ \emph {et~al.}(2015)\citenamefont {Loizu},
  \citenamefont {Hudson}, \citenamefont {Bhattacharjee}, \citenamefont
  {Lazerson},\ and\ \citenamefont {Helander}}]{LoizuHBLH2015}%
  \BibitemOpen
  \bibfield  {author} {\bibinfo {author} {\bibfnamefont {J.}~\bibnamefont
  {Loizu}}, \bibinfo {author} {\bibfnamefont {S.~R.}\ \bibnamefont {Hudson}},
  \bibinfo {author} {\bibfnamefont {A.}~\bibnamefont {Bhattacharjee}}, \bibinfo
  {author} {\bibfnamefont {S.}~\bibnamefont {Lazerson}}, \ and\ \bibinfo
  {author} {\bibfnamefont {P.}~\bibnamefont {Helander}},\ }\href {\doibase
  10.1063/1.4931094} {\bibfield  {journal} {\bibinfo  {journal} {Physics of
  Plasmas}\ }\textbf {\bibinfo {volume} {22}},\ \bibinfo {pages} {090704}
  (\bibinfo {year} {2015})}\BibitemShut {NoStop}%
\bibitem [{\citenamefont {Rosenbluth}\ \emph {et~al.}(1973)\citenamefont
  {Rosenbluth}, \citenamefont {Dagazian},\ and\ \citenamefont
  {Rutherford}}]{RosenbluthDR1973}%
  \BibitemOpen
  \bibfield  {author} {\bibinfo {author} {\bibfnamefont {M.~N.}\ \bibnamefont
  {Rosenbluth}}, \bibinfo {author} {\bibfnamefont {R.~Y.}\ \bibnamefont
  {Dagazian}}, \ and\ \bibinfo {author} {\bibfnamefont {P.~H.}\ \bibnamefont
  {Rutherford}},\ }\href@noop {} {\bibfield  {journal} {\bibinfo  {journal}
  {Phys. Fluids}\ }\textbf {\bibinfo {volume} {16}},\ \bibinfo {pages} {1894}
  (\bibinfo {year} {1973})}\BibitemShut {NoStop}%
\bibitem [{\citenamefont {Boozer}\ and\ \citenamefont
  {Pomphrey}(2010)}]{BoozerP2010}%
  \BibitemOpen
  \bibfield  {author} {\bibinfo {author} {\bibfnamefont {A.~H.}\ \bibnamefont
  {Boozer}}\ and\ \bibinfo {author} {\bibfnamefont {N.}~\bibnamefont
  {Pomphrey}},\ }\href@noop {} {\bibfield  {journal} {\bibinfo  {journal}
  {Physics of Plasmas}\ }\textbf {\bibinfo {volume} {17}},\ \bibinfo {pages}
  {110707} (\bibinfo {year} {2010})}\BibitemShut {NoStop}%
\bibitem [{\citenamefont {Loizu}\ and\ \citenamefont
  {Helander}(2017)}]{LoizuH2017}%
  \BibitemOpen
  \bibfield  {author} {\bibinfo {author} {\bibfnamefont {J.}~\bibnamefont
  {Loizu}}\ and\ \bibinfo {author} {\bibfnamefont {P.}~\bibnamefont
  {Helander}},\ }\href {\doibase 10.1063/1.4979678} {\bibfield  {journal}
  {\bibinfo  {journal} {Phys. Plasmas}\ }\textbf {\bibinfo {volume} {24}},\
  \bibinfo {pages} {040701} (\bibinfo {year} {2017})}\BibitemShut {NoStop}%
\bibitem [{\citenamefont {Berrut}\ and\ \citenamefont
  {Trefethen}(2004)}]{BerrutT2004}%
  \BibitemOpen
  \bibfield  {author} {\bibinfo {author} {\bibfnamefont {J.-P.}\ \bibnamefont
  {Berrut}}\ and\ \bibinfo {author} {\bibfnamefont {L.~N.}\ \bibnamefont
  {Trefethen}},\ }\href@noop {} {\bibfield  {journal} {\bibinfo  {journal}
  {SIAM Review}\ }\textbf {\bibinfo {volume} {46}},\ \bibinfo {pages} {501}
  (\bibinfo {year} {2004})}\BibitemShut {NoStop}%
\bibitem [{\citenamefont {Abramowitz}\ and\ \citenamefont
  {Stegun}(1972)}]{AbramowitzS1972}%
  \BibitemOpen
  \bibinfo {editor} {\bibfnamefont {M.}~\bibnamefont {Abramowitz}}\ and\
  \bibinfo {editor} {\bibfnamefont {I.~A.}\ \bibnamefont {Stegun}},\ eds.,\
  \href@noop {} {\emph {\bibinfo {title} {Handbook of Mathematical Functions
  with Formulas, Graphs, and Mathematical Tables}}},\ \bibinfo {edition}
  {10th}\ ed.\ (\bibinfo  {publisher} {National Bureau of Standards},\ \bibinfo
  {year} {1972})\BibitemShut {NoStop}%
\bibitem [{\citenamefont {Newcomb}(1962)}]{Newcomb1962}%
  \BibitemOpen
  \bibfield  {author} {\bibinfo {author} {\bibfnamefont {W.~A.}\ \bibnamefont
  {Newcomb}},\ }\href@noop {} {\bibfield  {journal} {\bibinfo  {journal}
  {Nuclear Fusion Supplement, Part 2}\ ,\ \bibinfo {pages} {451}} (\bibinfo
  {year} {1962})}\BibitemShut {NoStop}%
\bibitem [{\citenamefont {Zhou}\ \emph {et~al.}(2014)\citenamefont {Zhou},
  \citenamefont {Qin}, \citenamefont {Burby},\ and\ \citenamefont
  {Bhattacharjee}}]{ZhouQBB2014}%
  \BibitemOpen
  \bibfield  {author} {\bibinfo {author} {\bibfnamefont {Y.}~\bibnamefont
  {Zhou}}, \bibinfo {author} {\bibfnamefont {H.}~\bibnamefont {Qin}}, \bibinfo
  {author} {\bibfnamefont {J.~W.}\ \bibnamefont {Burby}}, \ and\ \bibinfo
  {author} {\bibfnamefont {A.}~\bibnamefont {Bhattacharjee}},\ }\href {\doibase
  10.1063/1.4897372} {\bibfield  {journal} {\bibinfo  {journal} {Phys.
  Plasmas}\ }\textbf {\bibinfo {volume} {21}},\ \bibinfo {pages} {102109}
  (\bibinfo {year} {2014})}\BibitemShut {NoStop}%
\bibitem [{\citenamefont {Pfefferl{\'{e}}}\ \emph {et~al.}(2020)\citenamefont
  {Pfefferl{\'{e}}}, \citenamefont {Noakes},\ and\ \citenamefont
  {Zhou}}]{PfefferleNZ2020}%
  \BibitemOpen
  \bibfield  {author} {\bibinfo {author} {\bibfnamefont {D.}~\bibnamefont
  {Pfefferl{\'{e}}}}, \bibinfo {author} {\bibfnamefont {L.}~\bibnamefont
  {Noakes}}, \ and\ \bibinfo {author} {\bibfnamefont {Y.}~\bibnamefont
  {Zhou}},\ }\href {\doibase 10.1088/1361-6587/ab8ca3} {\bibfield  {journal}
  {\bibinfo  {journal} {Plasma Physics and Controlled Fusion}\ }\textbf
  {\bibinfo {volume} {62}},\ \bibinfo {pages} {074004} (\bibinfo {year}
  {2020})}\BibitemShut {NoStop}%
\bibitem [{\citenamefont {Zhou}(2017)}]{Zhou2017}%
  \BibitemOpen
  \bibfield  {author} {\bibinfo {author} {\bibfnamefont {Y.}~\bibnamefont
  {Zhou}},\ }\emph {\bibinfo {title} {{Variational Integration for Ideal
  Magnetohydrodynamics and Formation of Current Singularities}}},\ \href
  {http://arxiv.org/abs/1708.08523} {Ph.D. thesis},\ \bibinfo  {school}
  {Princeton University} (\bibinfo {year} {2017}),\ \Eprint
  {http://arxiv.org/abs/1708.08523} {arXiv:1708.08523} \BibitemShut {NoStop}%
\bibitem [{\citenamefont {Lazerson}\ \emph {et~al.}(2016)\citenamefont
  {Lazerson}, \citenamefont {Loizu}, \citenamefont {Hirshman},\ and\
  \citenamefont {Hudson}}]{LazersonLHH2016}%
  \BibitemOpen
  \bibfield  {author} {\bibinfo {author} {\bibfnamefont {S.~A.}\ \bibnamefont
  {Lazerson}}, \bibinfo {author} {\bibfnamefont {J.}~\bibnamefont {Loizu}},
  \bibinfo {author} {\bibfnamefont {S.}~\bibnamefont {Hirshman}}, \ and\
  \bibinfo {author} {\bibfnamefont {S.~R.}\ \bibnamefont {Hudson}},\ }\href
  {\doibase 10.1063/1.4939881} {\bibfield  {journal} {\bibinfo  {journal}
  {Physics of Plasmas}\ }\textbf {\bibinfo {volume} {23}},\ \bibinfo {pages}
  {012507} (\bibinfo {year} {2016})}\BibitemShut {NoStop}%
\bibitem [{\citenamefont {Park}\ \emph {et~al.}(1980)\citenamefont {Park},
  \citenamefont {Monticello}, \citenamefont {White},\ and\ \citenamefont
  {Jardin}}]{ParkMWJ1980}%
  \BibitemOpen
  \bibfield  {author} {\bibinfo {author} {\bibfnamefont {W.}~\bibnamefont
  {Park}}, \bibinfo {author} {\bibfnamefont {D.~A.}\ \bibnamefont
  {Monticello}}, \bibinfo {author} {\bibfnamefont {R.~B.}\ \bibnamefont
  {White}}, \ and\ \bibinfo {author} {\bibfnamefont {S.~C.}\ \bibnamefont
  {Jardin}},\ }\href@noop {} {\bibfield  {journal} {\bibinfo  {journal} {Nucl.
  Fusion}\ }\textbf {\bibinfo {volume} {20}},\ \bibinfo {pages} {1181}
  (\bibinfo {year} {1980})}\BibitemShut {NoStop}%
\bibitem [{\citenamefont {Mikhailov}\ \emph {et~al.}(2019)\citenamefont
  {Mikhailov}, \citenamefont {Nührenberg},\ and\ \citenamefont
  {Zille}}]{MikhailovNZ2019}%
  \BibitemOpen
  \bibfield  {author} {\bibinfo {author} {\bibfnamefont {M.}~\bibnamefont
  {Mikhailov}}, \bibinfo {author} {\bibfnamefont {J.}~\bibnamefont
  {Nührenberg}}, \ and\ \bibinfo {author} {\bibfnamefont {R.}~\bibnamefont
  {Zille}},\ }\href {\doibase 10.1088/1741-4326/ab0f50} {\bibfield  {journal}
  {\bibinfo  {journal} {Nuclear Fusion}\ }\textbf {\bibinfo {volume} {59}},\
  \bibinfo {pages} {066002} (\bibinfo {year} {2019})}\BibitemShut {NoStop}%
\bibitem [{\citenamefont {Kim}\ \emph {et~al.}(2020)\citenamefont {Kim},
  \citenamefont {McFadden},\ and\ \citenamefont {Cerfon}}]{KimMC2020}%
  \BibitemOpen
  \bibfield  {author} {\bibinfo {author} {\bibfnamefont {E.}~\bibnamefont
  {Kim}}, \bibinfo {author} {\bibfnamefont {G.~B.}\ \bibnamefont {McFadden}}, \
  and\ \bibinfo {author} {\bibfnamefont {A.~J.}\ \bibnamefont {Cerfon}},\
  }\href {\doibase 10.1088/1361-6587/ab6d48} {\bibfield  {journal} {\bibinfo
  {journal} {Plasma Physics and Controlled Fusion}\ }\textbf {\bibinfo {volume}
  {62}},\ \bibinfo {pages} {044002} (\bibinfo {year} {2020})}\BibitemShut
  {NoStop}%
\bibitem [{\citenamefont {Cary}\ and\ \citenamefont
  {Hanson}(1991)}]{CaryH1991}%
  \BibitemOpen
  \bibfield  {author} {\bibinfo {author} {\bibfnamefont {J.~R.}\ \bibnamefont
  {Cary}}\ and\ \bibinfo {author} {\bibfnamefont {J.~D.}\ \bibnamefont
  {Hanson}},\ }\href {\doibase 10.1063/1.859829} {\bibfield  {journal}
  {\bibinfo  {journal} {Physics of Fluids B: Plasma Physics}\ }\textbf
  {\bibinfo {volume} {3}},\ \bibinfo {pages} {1006} (\bibinfo {year}
  {1991})}\BibitemShut {NoStop}%
\bibitem [{\citenamefont {Loizu}\ \emph {et~al.}(2020)\citenamefont {Loizu},
  \citenamefont {Huang}, \citenamefont {Hudson}, \citenamefont {Baillod},
  \citenamefont {Kumar},\ and\ \citenamefont {Qu}}]{LoizuHHBKQ2020}%
  \BibitemOpen
  \bibfield  {author} {\bibinfo {author} {\bibfnamefont {J.}~\bibnamefont
  {Loizu}}, \bibinfo {author} {\bibfnamefont {Y.-M.}\ \bibnamefont {Huang}},
  \bibinfo {author} {\bibfnamefont {S.~R.}\ \bibnamefont {Hudson}}, \bibinfo
  {author} {\bibfnamefont {A.}~\bibnamefont {Baillod}}, \bibinfo {author}
  {\bibfnamefont {A.}~\bibnamefont {Kumar}}, \ and\ \bibinfo {author}
  {\bibfnamefont {Z.~S.}\ \bibnamefont {Qu}},\ }\href {\doibase
  10.1063/5.0009110} {\bibfield  {journal} {\bibinfo  {journal} {Physics of
  Plasmas}\ }\textbf {\bibinfo {volume} {27}},\ \bibinfo {pages} {070701}
  (\bibinfo {year} {2020})}\BibitemShut {NoStop}%
\bibitem [{\citenamefont {Geraldini}\ \emph {et~al.}(2021)\citenamefont
  {Geraldini}, \citenamefont {Landreman},\ and\ \citenamefont
  {Paul}}]{GeraldiniLP2021}%
  \BibitemOpen
  \bibfield  {author} {\bibinfo {author} {\bibfnamefont {A.}~\bibnamefont
  {Geraldini}}, \bibinfo {author} {\bibfnamefont {M.}~\bibnamefont
  {Landreman}}, \ and\ \bibinfo {author} {\bibfnamefont {E.}~\bibnamefont
  {Paul}},\ }\href {\doibase 10.1017/s0022377821000428} {\bibfield  {journal}
  {\bibinfo  {journal} {Journal of Plasma Physics}\ }\textbf {\bibinfo {volume}
  {87}} (\bibinfo {year} {2021}),\ 10.1017/s0022377821000428}\BibitemShut
  {NoStop}%
\bibitem [{\citenamefont {Rodr{\'{\i}}guez}\ and\ \citenamefont
  {Bhattacharjee}(2021)}]{RodriguezB2021}%
  \BibitemOpen
  \bibfield  {author} {\bibinfo {author} {\bibfnamefont {E.}~\bibnamefont
  {Rodr{\'{\i}}guez}}\ and\ \bibinfo {author} {\bibfnamefont {A.}~\bibnamefont
  {Bhattacharjee}},\ }\href {\doibase 10.1063/5.0057186} {\bibfield  {journal}
  {\bibinfo  {journal} {Physics of Plasmas}\ }\textbf {\bibinfo {volume}
  {28}},\ \bibinfo {pages} {092506} (\bibinfo {year} {2021})}\BibitemShut
  {NoStop}%
\end{thebibliography}%

\end{document}